\documentclass[12pt]{article}

\usepackage{geometry}
\usepackage{amsmath, amssymb, amsthm}
\usepackage{mathrsfs}
\usepackage{bm, bbm}
\usepackage{amsfonts} 
\usepackage{dsfont}
\usepackage{verbatim}
\usepackage{titling}
\usepackage{txfonts}
\usepackage{url}
\usepackage{color}
\usepackage{graphicx}
\usepackage{datetime}
\usepackage{mathtools}
\usepackage{epstopdf}
\usepackage{float}
\usepackage{caption}
\usepackage{algorithmic}
\usepackage{algorithm}
\usepackage{array}
\usepackage{booktabs}
\usepackage[colorlinks,
linkcolor=blue,
anchorcolor=blue,
urlcolor=blue,
citecolor=blue]{hyperref}
\usepackage{verbatim}
\usepackage{threeparttable}
\usepackage{threeparttablex}
\usepackage{multirow}
\usepackage[round]{natbib}
\usepackage{listings}
\usepackage{rotating}
\usepackage{longtable}
\usepackage{cases}
 \newtheoremstyle{myexample}{5pt}{5pt}{\upshape}{}{\bfseries}{.}{ }{} \theoremstyle{myexample}

\newcommand{\D}{\mathcal{D}}

\def\T{{ \top }}

\newcommand{\cov}{{\rm cov}}
\newcommand{\var}{{\rm var}}
\newtheorem{condition}{Condition}[section]
\def\theequation{\arabic{section}.\arabic{equation}}
\newtheorem{theorem}{Theorem}[section]

\newtheorem{proposition}{Proposition}[section]

\newtheorem{lemma}{Lemma}[section]

\newtheorem{remark}{Remark}[section]

\newtheorem{example}{Example}[section]
\begin{document}

%\bibliographystyle{natbib}

% \def\spacingset#1{\renewcommand{\baselinestretch}%
% {#1}\small\normalsize} \spacingset{1}

% \allowdisplaybreaks[3] %allow align environment spread page

\title{\bf
Semiparametric efficient estimation of genetic relatedness with machine learning methods
%Double machine learning for genetic relatedness
}
\author{Xu Guo$^{1}$, Yiyuan Qian$^1$, Hongwei Shi$^1$, Weichao Yang$^1$, and Niwen Zhou$^2$\thanks{Corresponding Author: Niwen Zhou. All authors contributed equally to this work and are listed in the alphabetical order.}\\~\\
{\small \it $^{1}$ School of Statistics, Beijing Normal University, Beijing, China}\\
{\small \it $^{2}$ Center for Statistics and Data Science, Beijing Normal University, Zhuhai, China}
}

\date{}
%\date{\today}
\maketitle

\vspace{-0.5in}

\begin{abstract}
In this paper, we propose semiparametric efficient estimators of genetic relatedness between two traits in a model-free framework. Most existing methods require specifying certain parametric models involving the traits and genetic variants. However, the bias due to model misspecification may yield misleading statistical results. Moreover, the semiparametric efficient bounds for estimators of genetic relatedness are still lacking. In this paper, we develop semiparametric efficient estimators with machine learning methods and construct valid confidence intervals for two important measures of genetic relatedness: genetic covariance and genetic correlation, allowing both continuous and discrete responses. Based on the derived efficient influence functions of genetic relatedness, we propose a consistent estimator of the genetic covariance as long as one of genetic values is consistently estimated. The data of two traits may be collected from the same group or different groups of individuals. Various numerical studies are performed to illustrate our introduced procedures. We also apply proposed procedures to analyze Carworth Farms White mice genome-wide association study data.

\medskip

\noindent {\it Keywords:} Model misspecification; Genetic covariance; Semiparametric efficient bound; Confidence interval.

\end{abstract}

\section{Introduction}
Understanding genetic relatedness between complex traits is an important problem in human genetics research. In practical genetic studies, shared common genetic variants have been found in many complex diseases,
such as various autoimmune diseases \citep{zhernakova2009} and psychiatric disorders \citep{craddock2005}. Genetic relatedness analysis has a variety of downstream applications, which can help to find disease-associated genetic variation, improve polygenic risk prediction, and may contribute to improving nosology and diagnosis, risk stratification, and lifestyle interventions \citep{van2019genetic}.

Genetic covariance and genetic correlation are two popular measures of genetic relatedness.
Consider two responses $Y\in\mathbb{R}$ and $Z\in\mathbb{R}$,
such as complex traits, disease outcomes, or gene expressions, and $X\in\mathbb{R}^p$ being $p$-dimensional predictors, denoting as $p$ genetic variants. As introduced by \cite{van2019genetic} and \cite{wang2021unified}, the genetic covariance of  $Y$ and $Z$ can be defined as the covariance of their conditional mean functions
\begin{align*}
	I={\rm cov}\{m(X), h(X)\},
\end{align*}
where $m(X)=E(Y\mid X)$ and  $h(X)=E(Z\mid X)$ are the genetic values of $Y$ and $Z$, respectively. Subsequently, the genetic correlation can be defined as follows:
\begin{align*}
	\rho=\frac{{\rm cov}\{m(X),h(X)\}}{\sqrt{{\rm var}\{m(X)\}{\rm var}\{h(X)\}}}.
\end{align*}
Accordingly, $\rho$ is normalized as $-1\leq \rho\leq 1$, and thus it can be used to compare the genetic relatedness among multiple pairs.

In the genetic literature, methodological developments for estimating genetic relatedness are mainly based on family studies or genome-wide association studies (GWAS). Compared with traditional family-based approaches, GWAS-based methods do not require the studied phenotypes to be measured on the same individuals \citep{zhang2021comparison}. Thus it is promising in quantifying the overlapping genetic effects between pairs of traits based on GWAS data. Most of the GWAS-based methods are derived based on some specified regression models, such as linear mixed-effect model \citep{lee2012estimation,vattikuti2012heritability,yang2013polygenic}, linear fixed-effect model \citep{guo2019optimal}, and generalized linear model \citep{ma2022statistical, wang2021unified}.

However, the imposed stringent model structure assumptions are possibly not satisfied in practice, leading to biased estimators and inaccurate inference results. {Thus, it is important to develop estimation and inference procedures without any parametric model assumptions. To avoid model misspecification, flexible machine learning methods can be adopted to estimate the regression functions $m(X)$ and $h(X)$. A naive application would make the corresponding estimators inconsistent due to the complex and even black-box nature of machine learning methods. Besides, in many studies, the responses $Y$ and $Z$ are discrete. Most of the existing methods focus on continuous traits, and cannot be directly applied to estimate the genetic correlation between binary traits. \cite{weissbrod2018estimating} proposed a modified mixed effect model to deal with binary traits. Recently, \cite{ma2022statistical} and \cite{wang2021unified} also focus on estimation of genetic relatedness with binary traits and high dimensional GWAS data. However, these methods again would suffer from model misspecification. Moreover, even though many kinds of estimators have been proposed in the literature on genetic relatedness analysis, the semiparametric efficient bounds for estimators of genetic covariance and genetic correlation are still not established. All these critical issues require new methodologies and theoretical results.}

Aiming to address the above challenges and questions, this paper proposes efficient and model-free estimators of genetic covariance and genetic correlation for both continuous and discrete traits. The main contributions of this paper are summarized as follows.
Firstly, we derive the efficient influence functions of genetic covariance and genetic correlation, which provides the semiparametric efficient bounds for estimators of genetic covariance and genetic correlation.
Secondly, based on the efficient influence function, we propose a consistent estimator of the genetic covariance as long as either $m(X)$ or $h(X)$ is consistently estimated.
Thirdly, to guarantee valid statistical inference for genetic covariance and genetic correlation with possible high-dimensional GWAS data, we estimate genetic covariance and genetic correlation by combining the results of efficient influence functions and the sample-splitting strategy. The proposed estimators are model-free and semiparametric efficient.
Fourthly, our methods are very general. Actually, our procedures are applicable to the data of two continuous or discrete traits with overlap or without overlap samples.

The paper is organized as follows. Section \ref{section:sec2} presents the estimation and inference procedures of genetic covariance. In Section \ref{section:sec3}, these procedures are also established for genetic correlation. Further, the case with the discrete response is also
considered in Section \ref{section:sec4}.
%To better investigate genetic relatedness, a local detection procedure is proposed in section 5.
The numerical performance of the proposed methods is presented in Section \ref{section:sec5}.  In Section \ref{section:sec6}, we present a real data example. Conclusions and discussions are given in Section \ref{section:sec7}.
The proofs of theoretical results are given in the Appendix.

\section{Estimation for genetic covariance}\label{section:sec2}
In this section, we propose estimators for the genetic covariance $I$. Now suppose that $\D_y=\{(X_i, Y_i),i=1,\ldots,N_y\}$ and $\D_z=\{(X_j, Z_j),j=1,\ldots,N_z\}$ are two independent and identically distributed random samples from the population $(X,Y)$ and $(X,Z)$, respectively. Let $\D_0=\D_y\cap\D_z$, $\D=\D_y\cup\D_z$ and $N_0=|\D_0|, N=|\D|=N_y+N_z-N_0$.
{In this paper, we allow the data of two traits to be collected from the same group or different groups of individuals. Specifically, when the data are from two independent samples with $N_0=0$, we call it non-overlap. While  the data are from two samples with
	$N_0>0$, we call it overlap. In particular, if the data are from the same samples with $N_0=N_y=N_z$, we call it fully overlap.
}
%In this paper, we allow the data of two traits be collected from the same group or different groups of individuals. When the data are from two independent samples, $N_0=0$. While the data are from the same samples, $N_0=N_y=N_z$.
Suppose that $\hat m(X)$ and $\hat h(X)$ are two suitable estimators of $m(X)$ and $h(X)$, respectively. Recall that $I=E\{m(X)h(X)\}-E(Y)E(Z)$, one may consider the following natural plug-in estimator of $I$:
\begin{align*}
	I_N= N^{-1}\sum_{i\in\D}\hat m(X_i)\hat h(X_i)-\bar Y_N\bar Z_N.
\end{align*}
Here $\bar{Y}_N=N_{y}^{-1}\sum_{i=1}^{N_y}Y_i, \bar{Z}_N=N_{z}^{-1}\sum_{i=1}^{N_z}Z_i$. However, this natural estimator cannot even be consistent due to the overfitting and bias of the first term $N^{-1}\sum_{i\in\D}\hat m(X_i)\hat h(X_i)$ when we adopt flexible machine learning methods to obtain $\hat m(X)$ and $\hat h(X)$.

To obtain consistent and asymptotic normal estimators, it is shown that the efficient influence function (EIF) plays a critical role.
Before this, we give some notation. Recall that the observation data is $\D=\D_y\cup\D_z$ and some units may not be observed in both samples. Let $T_{y,i}=1$ if $Y_i$ is observed, otherwise $T_{y,i}=0$. The definition of $T_{z,i}$ is similar. Note that whether we can observe $Y$ or $Z$ are unrelated to the data. Thus, the missingness of $Y$ or $Z$ is completely at random, i.e. $T_y\perp (Y,X)$ and $T_z\perp (Z, X)$. By adopting the notation of missing data literature, let $Y^*_i=T_{y,i}Y_i$ and $Z^*_i=T_{z,i}Z_i$. See \cite{tsiatis2007semiparametric} for further reference. Thus, the observation data is $\{O_i\}_{i=1}^N=\{X_i,Y^*_i,Z^*_i,T_{y,i},T_{z,i}\}_{i=1}^N$.
%Now we first derive the efficient influence function for $I$ in the following theorem.
To derive the efficient influence function for $I$, the following regular condition is required.
\begin{condition}
	\label{condition1}
	$E\{\sigma^4(X)\}$, $E\{\delta^4(X)\}$, $E[\{m(X)-E(Y)\}^4]$, and $E[\{h(X)-E(Z)\}^4]$ are finite.
\end{condition}
Here $\sigma^2(X)=E(\epsilon^2\mid X)$ and $\delta^2(X)=E(\eta^2\mid X)$ with $\epsilon=Y-m(X), \eta=Z-h(X)$.
This condition is very mild and it guarantees that the variance of EIF for $I$ is finite.

\begin{theorem}\label{theorem_eif}
	Under the Condition \ref{condition1},  the efficient influence function for $I$ is given by
	\begin{align}\label{eif1}
		S_{yz}=\ &\frac{\mathbb{I}(T_y = 1)}{\Pr(T_y=1)} \{Y^*-E(Y^*\mid X,T_y=1)\}\{E(Z^*\mid X,T_z=1)-E(Z^*\mid T_z=1)\}\notag\\
		&+\frac{\mathbb{I}(T_z = 1)}{\Pr(T_z=1)} \{Z^*-E(Z^*\mid X,T_z=1)\}\{E(Y^*\mid X,T_y=1)-E(Y^*\mid T_y=1)\}\notag\\
		&+\{E(Y^*\mid X,T_y=1)-E(Y^*\mid T_y=1)\}\{E(Z^*\mid X,T_z=1)-E(Z^*\mid T_z=1)\}-I,
	\end{align}
	where $\mathbb{I}(\cdot)$ is the indicator function.
	{Hence, the semiparametric efficiency bound for $I$ is  equal to
		\begin{align*}
			\var(S_{yz})=&\ \frac{1}{\Pr(T_y=1)} E[\sigma^2(X)\{h(X)-E(Z)\}^2]+\frac{1}{\Pr(T_z=1)} E[\delta^2(X)\{m(X)-E(Y)\}^2]\nonumber\\
			&+E([\{m(X)-E(Y)\}\{h(X)-E(Z)\} -I]^2)\nonumber\\
			&+\frac{\Pr(T_y=1,T_z=1)}{\Pr(T_y=1)\Pr(T_z=1)} E[\epsilon\eta\{m(X)-E(Y)\}\{h(X)-E(Z)\}].
		\end{align*}
	}
	%Hence, the semiparametric efficiency bound for $I$ is  equal to $\mbox{Var}(S_{yz}).$
\end{theorem}
{\begin{remark}
		Under the non-overlap case, the semiparametric efficiency bound for $I$ is \begin{align*}
			\var(S_{yz}) =&\  \frac{1}{\Pr(T_y=1)}E[\sigma^2(X)\{h(X)-E(Z)\}^2] +\frac{1}{\Pr(T_z=1)} E[\delta^2(X)\{m(X)-E(Y)\}^2]\nonumber\\
			&+E([\{m(X)-E(Y)\}\{h(X)-E(Z)\} -I]^2).
		\end{align*}
		While under the fully overlap case, the semiparametric efficiency bound for $I$ is \begin{align*}
			\var(S_{yz})=&\  E[\sigma^2(X)\{h(X)-E(Z)\}^2] + E[\delta^2(X)\{m(X)-E(Y)\}^2]\nonumber\\
			&+E([\{m(X)-E(Y)\}\{h(X)-E(Z)\} -I]^2)\nonumber\\
			&+E[\epsilon\eta\{m(X)-E(Y)\}\{h(X)-E(Z)\}].
		\end{align*}
	\end{remark}
}

From the above efficient influence function for $I$, a consistent estimator $\tilde {I}_N$ for $I$ can be delivered by making the sample average of the estimated influence function zero, i.e.
\begin{align*}
	\tilde I_N=&N_y^{-1}\sum_{i=1}^{N_y}\{Y_i-\hat m(X_i)\}\{\hat h(X_i)-\bar Z_N\}+ N_z^{-1}\sum_{i=1}^{N_z}\{\hat m(X_i)-\bar Y_N\}\{Z_i-\hat h(X_i)\}\\
	&+ N^{-1}\sum_{i=1}^{N}\{\hat m(X_i)-\bar Y_N\}\{\hat h(X_i)-\bar Z_N\}.
\end{align*}

Suppose that $\hat m(X_i)$ and $\hat h(X_i)$ are consistent estimators of $\tilde m(X_i)$ and $\tilde h(X_i)$ in the sense that $E[\{\hat m(X_i)-\tilde m(X_i)\}^2]=o(1)$ and $E[\{\hat h(X_i)-\tilde h(X_i)\}^2]=o(1)$, while  $\tilde m(X_i)$ and $\tilde h(X_i)$  may not be equal to $m(X_i)$ and $h(X_i)$, respectively. In the following theorem, we show the consistency of
$\tilde I_N$ under mild conditions.

\begin{theorem}\label{theorem_con}
	Under the Condition \ref{condition1}, $E\{\tilde m^2(X)\}$ and $E\{\tilde h^2(X)\}$ are finite, and further
	$E[\{\tilde m(X)-m(X)\}\{h(X)-\tilde h(X)\}]=0$, $\tilde I_N$ is a consistent estimator of $I$.
\end{theorem}

%Under condition that $N_0/N_y\rightarrow 1$ and $N_0/N_z\rightarrow 1$, we can derive that $N=N_y(1+o(1))=N_z(1+o(1))$. This asks the non-overlap part is relatively negligible. This may be restrictive. We will explore the situation that the two outcomes collected separately in the numerical studies.

Theorem \ref{theorem_con} presents very interesting results. It shows that the above $\tilde I_N$ is consistent in general situation. Even though both $\hat m(X_i)$ and $\hat h(X_i)$ are completely inconsistent estimators of $m(X_i)$ and $h(X_i)$, the above $\tilde I_N$ can still be consistent estimator of $I$ given $\tilde m(X)-m(X)$ and $h(X)-\tilde h(X)$ are uncorrelated. If either $\hat m(X_i)$ or $\hat h(X_i)$, not necessary both, is consistent estimator of $m(X_i)$ or $h(X_i)$, then the $\tilde I_N$ is a consistent estimator of the genetic covariance $I$.

However, we note that for making statistical inference, the above $\tilde I_N$ is not enough. In fact, terms such as
$N_y^{-1}\sum_{i=1}^{N_y}\epsilon_i\{\hat h(X_i)-\tilde h(X_i)\}$ are not $o_p(N^{-1/2})$. To this end, sample splitting is required. To be precise, we split the whole data set $\D_y$ randomly into two independent data sets $\D_{1y}$ and $\D_{2y}$. Let $|\D_{2y}|=|\D_{1y}|=n_y$. Similarly $\D_z$ is divided into $\D_{1z}$ and $\D_{2z}$ with $|\D_{2z}|=|\D_{1z}|=n_z$. Denote $\D_{0k}=\D_{ky}\cap\D_{kz}$, $\D_k=\D_{ky}\cup\D_{kz}$, $n_0=|\D_{0k}|$, $n=|\D_k|=n_y+n_z-n_0$, $k=1,2$. Denote the estimators of $m(X)$ and $h(X)$ based on $\D_{1}$ as $\hat m_{\D_{1}}(X)$ and $\hat h_{\D_{1}}(X)$, respectively. Consider
\begin{align}\label{est_in}
%	\begin{aligned}
		I_n=&\ n_y^{-1}\sum_{i\in\D_{2y}}\{Y_i-\hat m_{\D_{1}}(X_i)\}\{\hat h_{\D_1}(X_i)-\bar Z_n\}+ n_z^{-1}\sum_{i\in\D_{2z}}\{\hat m_{\D_1}(X_i)-\bar Y_n\}\{Z_i- \hat h_{\D_1}(X_i)\}\notag\\
		&+ n^{-1}\sum_{i\in\D_{2}}\{\hat m_{\D_1}(X_i)-\bar Y_n\}\{\hat h_{\D_1}(X_i)-\bar Z_n\}.	
%	\end{aligned}
\end{align}
Here $\bar{Y}_n=n_y^{-1}\sum_{i\in\D_{2y}}Y_i, \bar{Z}_n=n_z^{-1}\sum_{i\in\D_{2z}}Z_i$.

The following conditions are required for theoretical analysis.
\begin{condition}
	\label{condition2}
	$E[\{\hat m_{\D_1}(X)-m(X)\}^2]=o(n_y^{-1/2})$ and $E[\{\hat h_{\D_1}(X)-h(X)\}^2]=o(n_z^{-1/2})$.
\end{condition}
\begin{condition}
	\label{condition3}
	$E[\sigma^2(X)\{\hat h_{\D_1}(X)-h(X)\}^2]=o(1)$ and $E[\delta^2(X)\{\hat m_{\D_1}(X)-m(X)\}^2]=o(1)$.
\end{condition}
\begin{condition}
	\label{condition4}
	$E[\{m(X)-E(Y)\}^2\{\hat h_{\D_1}(X)-h(X)\}^2]=o(1)$ , $E[\{h(X)-E(Z)\}^2\{\hat m_{\D_1}(X)-m(X)\}^2]=o(1)$, and $E[\{\hat m_{\D_1}(X)-m(X)\}^2\{\hat h_{\D_1}(X)-h(X)\}^2]=o(1)$.
\end{condition}

The above conditions are mild about moments and estimation errors. The Condition \ref{condition2} is fairly general, and has been commonly adopted in the literature. See for instance \cite{chernozhukov2018double} and \cite{vansteelandt2020assumption}. When both $\sigma^2(X)$ and $\delta^2(X)$ are finite, Condition \ref{condition3} holds directly under Condition \ref{condition2}. Both Conditions \ref{condition3} and \ref{condition4} are required to control properly the quadratic term and beyond in asymptotic analysis.

Further we define $I_n^{*}=(I_n+I'_n)/2$,  where $I'_n$ is similarly defined by swapping the role of $ \D_{1}$ and $ \D_{2}$. Since $I_n$ and $I'_n$ are asymptotically independent, the following theorem provides a solution for constructing confidence intervals for the genetic covariance.

\begin{theorem}\label{theorem_asy1}
	Suppose Conditions \ref{condition1}--\ref{condition4} are satisfied, we have
	\begin{align*}
		I^*_n-I=\sum_{i\in\D}S^*_{yz,i}+o_p( N_z^{-1/2}+N_y^{-1/2}).
	\end{align*}
	Here
	\begin{align*}
		S^*_{yz,i}=N_y^{-1}\epsilon_i^*\mathbb{I}(T_{y,i}=1) + N_z^{-1}\eta_i^*\mathbb{I}(T_{z,i}=1)+ N^{-1}\xi^*.
	\end{align*}
	with $\epsilon_i^*=\epsilon_i\{h(X_i)-E(Z)\}, \eta_i^*=\eta_i\{m(X_i)-E(Y)\}$ and $\xi_i^*=\{m(X_i)-E(Y) \}\{h(X_i)-E(Z)\}-I$. Further, the asymptotic normality of $I^*_n$ is
	\begin{align*}
		\sigma^{-1}(I_n^*-I) \overset{d}{\rightarrow} N(0,1).
	\end{align*}
	Here $
		\sigma^2= N_y^{-1}E(\epsilon^{*2})+ N_z^{-1}E(\eta^{*2})+ N^{-1}E(\xi^{*2}) + (N_yN_z)^{-1}N_{0}
		E(\epsilon^*\eta^*). $
\end{theorem}

\begin{remark}
	Suppose that
	\begin{align*}
		\Pr(T_y=1)=\lim_{(N,N_y)\rightarrow \infty}N_y/N,~
		\Pr(T_z=1)=\lim_{(N,N_z)\rightarrow \infty}N_z/N, ~
		\mbox{and }\Pr(T_y=1,T_z=1)=\lim_{(N, N_0) \rightarrow \infty}N_0/N.
	\end{align*}
	We can obtain that $N\sigma^2\rightarrow
	\var(S_{yz})$, which implies that the proposed estimator $I_n^*$ is semiparametric efficient in the sense that its asymptotic variance achieves the semiparametric efficient bound.
\end{remark}

In the above theorem, $\sigma^2$ is generally unknown and should be estimated. To this aim, denote
\begin{align*}
	\hat\delta_i=&\ {n_y}^{-1}\{Y_i-\hat m_{\D_{1}}(X_i)\}\{\hat h_{\D_1}(X_i)-\bar Z_n\}\mathbb{I}(T_{y,i}=1) +{n_z}^{-1}\{\hat m_{\D_1}(X_i)-\bar Y_n\}\{Z_i- \hat h_{\D_1}(X_i)\}\mathbb{I}(T_{z,i}=1)\\
	&+ n^{-1}\{\hat m_{\D_1}(X_i)-\bar Y_n\}\{ \hat h_{\D_1}(X_i)-\bar Z_n\}.
\end{align*}
Then $I_n=\sum_{i\in\D_2}\hat\delta_i$. A variance estimator for $I_n$ is
\begin{align*}
	K_1=\sum_{i\in\D_2}(\hat\delta_i-n^{-1}I_n)^2.
\end{align*}
$K_2$ is similarly defined by swapping the role of $\D_1$ and $\D_2$. Let $\hat\sigma^2=(K_1+K_2)/4$. It is shown that $\hat\sigma/\sigma\overset{p}{\rightarrow} 1$. Hence, an efficient confidence interval at the confidence level of $1-\alpha$ for $I$ is
\begin{align}
	\mbox{CI}(\alpha)=\left( I_n^*-z_{\alpha/2}\cdot\hat\sigma,\,\,
	I_n^*+z_{\alpha/2}\cdot\hat\sigma \right).
\end{align}
The following theorem establishes the validity of the above confidence interval for $I$.
\begin{theorem}
	\label{theorem_ci}
	Suppose  $E(\epsilon_i^{*4})$, $E(\eta_i^{*4})$ and $E(\xi_i^{*4})$ are finite, and Conditions \ref{condition2}--\ref{condition4}  are satisfied. We have
	\begin{align*}
		\hat{\sigma}^{-1}(I_n^*-I) \overset{d}{\rightarrow} N(0,1),
	\end{align*}
	and
	\begin{align*}
		\lim_{(N_y,N_z)\rightarrow\infty}\Pr\left\{I\in \mbox{CI}(\alpha)\right\}=1-\alpha.
	\end{align*}
\end{theorem}

\section{Estimation for genetic correlation}\label{section:sec3}

In this section, we consider the inference of genetic correlation which is a standardization of genetic covariance, defined as
\begin{align*}
	\rho=\frac{\cov\{m(X),h(X)\}}{\sqrt{\var\{m(X)\}\var\{h(X)\}}},
\end{align*}
where $m(X)=E(Y\mid X)$ and  $h(X)=E(Z\mid X)$.
To simplify the presentation, we give some notations here.  Denote $B_0^y=\var\{m(X)\},~B_0^z=\var\{h(X)\}$. Further let
\begin{align*}
	S_{yy} &= 2\frac{\mathbb{I}( {T}_y = 1)}{\Pr(T_y=1)} \{Y^*-m(X)\}\{m(X)-E(Y)\}+\{m(X)-E(Y)\}^2-B_0^y,\\
	S_{zz} &= 2\frac{\mathbb{I}({T}_z = 1)}{\Pr(T_z=1)} \{Z^*-h(X)\}\{h(X)-E(Z)\}+\{h(X)-E(Z)\}^2-B_0^z,
\end{align*}
which are the efficient influence functions for $B_0^y$ and $B_0^z$ respectively by Theorem~\ref{theorem_eif}.
We then obtain the efficient influence function of $\rho$ in the following Proposition.

\begin{proposition}\label{theorem_eifp}
	Under the Condition \ref{condition1}, the efficient influence function for $\rho$ is
	\begin{align}\label{eif2}
		\phi_1(O,\mathcal{P})
		&=\frac{S_{yz}}{\sqrt{B_0^yB_0^z}}-\rho\frac{S_{yy}}{2B_0^y}-\rho\frac{S_{zz}}{2B_0^z}.
	\end{align}
\end{proposition}

Motivated by the efficient influence functions for $B_0^y$ and $B_0^z$, we consider the following  estimators:
\begin{align*}
	\hat B_{0}^y &= n_y^{-1}\sum_{i\in\D_{2y}}2\{Y_i-\hat m_{\D_{1}}(X_i)\}\{\hat m_{\D_1}(X_i)-\bar Y_n\}+n^{-1}\sum_{i\in\D_{2}}\{\hat m_{\D_1}(X_i)-\bar Y_n\}^2,\\
	\hat B_{0}^z &= n_z^{-1}\sum_{i\in\D_{2z}}2\{Z_i- \hat h_{\D_1}(X_i)\}\{\hat h_{\D_1}(X_i)-\bar Z_n\}+ n^{-1}\sum_{i\in\D_{2}}\{\hat h_{\D_1}(X_i)-\bar Z_n\}^2.
\end{align*}

Further, we can get a consistent estimator $\rho_n$ for $\rho$  by making the sample average of the estimated influence functions zero, i.e.
\begin{align}
	\rho_{n}=\frac{I_{n}}{\sqrt{\hat B_{0}^y\hat B_{0}^z}}.
\end{align}
Further we define $\rho_n^{*}=(\rho_n+\rho'_n)/2$,  where $\rho'_n$ is similarly defined by swapping the role of $\D_1$ and $\D_2$. The asymptotic properties for $\rho_n^*$ are stated in the following theorem.

\begin{theorem}\label{theorem_asyp}
	Suppose Conditions \ref{condition1}--\ref{condition4} are satisfied. We have
	\begin{align}
		\rho_n^*-\rho = \sum_{i\in\D}\phi^*_1(O_i)+ o_p(N_z^{-1/2}+N_y^{-1/2}).
	\end{align}
	Here
	\begin{align*}
		\phi_1^*(O_i)=\frac{S^*_{yz,i}}{\sqrt{B_0^yB_0^z}}-\rho\frac{S^*_{yy,i}}{2B_0^y}-\rho\frac{S^*_{zz,i}}{2B_0^z}
	\end{align*}
	with
	\begin{align*}
		S^*_{yy,i} &= 2\epsilon_{y,i}^*\mathbb{I}(T_{y,i}=1)+\xi_{y,i}^*, \text{ and } S^*_{zz,i} = 2\eta_{y,i}^*\mathbb{I}(T_{z,i}=1)+\xi_{z,i}^*,
	\end{align*}
	where $\epsilon_{y,i}^*=\epsilon_i\{ m(X_i)-E(Y)\}$, $\xi_{y,i}^*=\{m(X_i)-E(Y)\}^2-B_0^y$. $\eta_{y,i}^*$ and $\xi_{z,i}^*$  are similarly defined. And the asymptotic normality of $\rho^*_n$ is
	\begin{align}
		\sigma_r^{-1}(\rho_n^*-\rho) \overset{d}{\rightarrow} N(0,1).
	\end{align}
	Here $\sigma_r^2=\var\{\sum_{i\in\D}\phi^*_1(O_i)\}.$
\end{theorem}

Based on Theorem \ref{theorem_asyp} and Proposition \ref{theorem_eifp}, we can see that the proposed estimator for $\rho$ is semiparametric efficient. Further, we can similarly construct the estimator for $\sigma_r^2$. Define
\begin{align*}
	J_1&=\sum_{i\in \D_2}\biggl(\frac{\hat S_{yz,i}}{\sqrt{\hat B_0^y\hat B_0^z}}-\rho_n\frac{\hat S_{yy,i}}{2\hat B_0^y}-\rho_n\frac{\hat S_{zz,i}}{2\hat B_0^z}\biggr)^2,
\end{align*}
where
\begin{align*}
	\hat S_{yz,i}=&\ n_y^{-1}\{Y_i-\hat m_{\D_{1}}(X_i)\}\{\hat h_{\D_1}(X_i)-\bar Z_n\}\mathbb{I}(T_{y,i}=1)\\
	&+n_z^{-1}\{\hat m_{\D_1}(X_i)-\bar Y_n\}\{Z_i- \hat h_{\D_1}(X_i)\}\mathbb{I}(T_{z,i}=1)\\
	&+n^{-1}[\{\hat m_{\D_1}(X_i)-\bar Y_n\}\{ \hat h_{\D_1}(X_i)-\bar Z_n\}-I_n],
\end{align*}
and $\hat S_{yy,i}$ and $\hat S_{zz,i}$ are similarly defined.
%Here for $i\in\D_2$, $\hat\epsilon_i=Y_i-\hat m_{\D_1}(X_i), \hat\eta_i=Z_i-\hat h_{\D_1}(X_i), \overline{m(X)h(X)}
%=n^{-1}\sum_{i\in\D_2}\hat m_{\D_1}(X_i)\hat h_{\D_1}(X_i)$.
$J_2$ is similarly defined by swapping the role of $\D_1$ and $\D_2$. Let $\hat \sigma_r^2=(J_1+J_2)/4$. Hence, a valid confidence interval at the confidence level of $1-\alpha$ for $\rho$ is
\begin{align}
	\mbox{CI}_r(\alpha)=(\rho_n^*-z_{\alpha/2}\cdot\hat \sigma_r,\ \rho_n^*+z_{\alpha/2}\cdot\hat \sigma_r).
\end{align}
Similarly, as Theorem \ref{theorem_ci}, the validity of the proposed confidence interval can be verified. To save space, the details of the proof are omitted but are available on request from the authors.

%\iffalse
%\begin{theorem}\label{theorem_cir}
%	Suppose  $E(\epsilon_i^{*4})$, $E(\eta_i^{*4})$, $E(\xi_i^{*4})$, $E(\epsilon_{y,i}^{*4})$, and $E(\eta_{z,i}^{*4})$ are finite, $E[\{\hat m_{\D_1}(X)-m(X)\}^4]=o(n_y^{-1})$ and $E[\{\hat h_{\D_1}(X)-h(X)\}^4]=o(n_z^{-1})$, and Conditions C1-C4 are satisfied.   We have
%	\begin{align}
%		\frac{\rho_n^*-\rho }{\hat \sigma_r} \overset{d}{\rightarrow} N\Big(0,1\Big).
%	\end{align}
%	So that
%	\begin{align}
%		&\lim_{N_y,N_z\rightarrow\infty}\Pr(\rho\in  \mbox{CI}_r(\alpha))=1-\alpha.
%	\end{align}
%	%Here $$S=[Y-E(Y|X)][E(Z|X)-E(Z)]+[E(Y|X)-E(Y)][Z-E(Z)].$$
%\end{theorem}
%\fi

\section{Generalized genetic covariance and generalized genetic correlation}\label{section:sec4}

In the above sections, we focus on continuous responses $Y$ and $Z$. In some applications, the interested response $Y$ or $Z$ may be discrete.
Thus in this section, we consider generalized genetic covariance and generalized genetic correlation which allow the outcomes to be different types.
Recall that  $m(X)=E(Y\mid X)$ and $h(X)=E(Z\mid X)$ are the genetic values of $Y$ and $Z$, respectively.
In such situations, with some abuse of notation, we define
\begin{align*}
	& I=\cov[g_1\{m(X)\}, g_2\{h(X)\}],\quad \text{and}\quad \rho=\frac{\cov[g_1\{m(X)\}, g_2\{h(X)\}]}{\sqrt{\var[g_1\{m(X)\}]\var[g_2\{h(X)\}]}}.
\end{align*}
Here $g_1(\cdot)$ and $g_2(\cdot)$ are two known link functions. For instance, for a binary response, the link function can be taken as $
	g_1(x)=\log x/(1-x). $

To simplify the illustration, let $E(g_1) = E[g_1\{m(X)\}]$ and $E(g_2)=E[g_2\{h(X)\}]$.
To derive the efficient influence function for $I$, the following regular conditions are required.
\begin{condition}
	\label{condition5}
	The functions $g_1(\cdot)$ and $g_2(\cdot)$ are  continuous differentiable with $|g_i'(\cdot)|\leq C<\infty$, $i=1,2$. Further, $g'_i(\cdot),\,i=1,2$ satisfy the Lipschitz condition for a positive constant $L>0$,
	\begin{align*}
		|g'_i(x_1)-g'_i(x_2)|\leq L|x_1-x_2|\quad \text{for all $x_1,x_2\in\mathbb{R}$.}
	\end{align*}
\end{condition}
\begin{condition}
	\label{condition6}
	$E\{\sigma^4(X)\}$, $E\{\delta^4(X)\}$, $E([g_1\{m(X)\}-E(g_1)]^4)$ and $E([g_2\{h(X)\}-E(g_2)]^4)$  are finite.
\end{condition}
Recall that $\sigma^2(X)=E(\epsilon^2\mid X)$ and $\delta^2(X)=E(\eta^2\mid X)$ with $\epsilon=Y-m(X), \eta=Z-h(X)$. Condition \ref{condition5} is mild and regular. The link functions of common models satisfy this condition, including the standard linear model, logistic model and multinomial logistic model.
\cite{wang2021unified} also requires this regular condition on the link function. Condition \ref{condition6} plays the same role as Condition \ref{condition1} to ensure the variance of EIF for $I$ is finite.

\begin{proposition}\label{theorem_eif2}
	Under the Conditions  \ref{condition5} and \ref{condition6}, we have\\
		{\rm (a)} The efficient influence function for $I$ is given by
		\begin{align*}
			S_{yz}=&\ \frac{\mathbb{I}({T}_y = 1)}{\Pr(T_y=1)}g'_1\{m(X)\}\{Y^*-m(X)\}[g_2\{h(X)\}-E(g_2)] \\
			&+\frac{\mathbb{I}({T}_z = 1)}{\Pr(T_z=1)}g'_2\{h(X)\}\{Z^*-h(X)\} [g_1\{m(X)\}-E(g_1)] \\
			&+[g_1\{m(X)\}-E(g_1)][g_2\{h(X)\}-E(g_2)]-I.
		\end{align*}
		{\rm (b)} The efficient influence function for $\rho$ is
		\begin{align*}
			\phi_1(O,\mathcal{P}) = \frac{S_{yz}}{\sqrt{B_0^yB_0^z}}-\rho\frac{S_{yy}}{2B_0^y}-\rho\frac{S_{zz}}{2B_0^z},
		\end{align*}
		where
		\begin{align*}
			S_{yy}&= 2\frac{\mathbb{I}({T}_y = 1)}{\Pr(T_y=1)}g'_1\{m(X)\}\{Y^*-m (X)\}[g_1\{m(X)\}-E(g_1)] + [g_1\{m(X)\}-E(g_1)]^2-B_0^y,\\
			S_{zz}&= 2\frac{\mathbb{I}({T}_z = 1)}{\Pr(T_z=1)}g'_2\{h(X)\}\{Z^*-h(X)\}[g_2\{h(X)\}-E(g_2)] + [g_2\{h(X)\}-E(g_2)]^2-B_0^z.
		\end{align*}
	As a result, the semiparametric efficiency bound for $I$ and $\rho$ are equal to $\var(S_{yz})$ and $\var\{\phi_1(O,\mathcal{P})\}$, respectively.
\end{proposition}

The expression of EIF varies across different conditions. Here are two examples.
\begin{example}
	When $g_1(x)=g_2(x)=x$, the above EIFs for $I$ and $\rho$ are coincide with  \eqref{eif1} and \eqref{eif2} respectively.
\end{example}
\begin{example}
	When $Y$ and $Z$ are binary from the logistic regression models, i.e.
\begin{align*}
		g_1\{m(X)\}=\log\biggl\{\frac{m(X)}{1-m(X)}\biggr\}=X^\T \beta, \text{ and }g_2\{h(X)\}&=\log\biggl\{\frac{h(X)}{1-h(X)}\biggr\}=X^\T \gamma,
\end{align*}
	which implies that $m(X)={\exp(X^\T \beta)}/\{1+\exp(X^\T \beta)\}$,
	$h(X)={\exp(X^\T \gamma)}/\{1+\exp(X^\T \gamma)\}$, and $I=\beta^\T\Sigma \gamma$ with $\Sigma=\var(X)$. It follows that the efficient influence function for $I$ is
	\begin{align*}
		S_{yz}=&\ \frac{\mathbb{I}({T}_y = 1)}{\Pr(T_y=1)}\frac{1}{m(X)\{1-m(X)\}}\{Y^*-m(X)\}\{X-E(X)\}^\T \gamma\\
		&+\frac{\mathbb{I}({T}_z = 1)}{\Pr(T_z=1)}\frac{1}{h(X)\{1-h(X)\}}\{Z^*-h(X)\} \{X-E(X)\}^\T \beta  \\
		&+\beta^\T\{X-E(X)\}\{X-E(X)\}^\T \gamma-\beta^\T\Sigma \gamma.
	\end{align*}
\end{example}

Based on the above EIF, we consider the following estimator for $I$,
\begin{align*}
	I_n=&\ n_y^{-1}\sum_{i\in\D_{2y}}
	\hat g'_{1i}(Y_i-\hat m_i)(\hat g_{2i}-\bar{g}_2)+ n_z^{-1}\sum_{i\in\D_{2z}}\{\hat{g}_{2i}'(Z_i-\hat h_i)\} (\hat g_{1i}-\bar{g}_1)\\
	&+ n^{-1}\sum_{i\in\D_2}(\hat g_{2i}-\bar{g}_2)(\hat g_{1i}-\bar{g}_1).
\end{align*}
Here $\hat m_i=\hat m_{\D_1}(X_i)$, $ \hat g_{1i}=g_1(\hat m_i)$, $\hat g'_{1i}=g_1'(\hat m_i)$, $\bar{g}_1=n^{-1}\sum_{i\in\D_2}\hat g_{1i}$.The notations of $\hat h_i$, $\hat g_{2i}$, $\hat g'_{2i}$ and $\bar{g}_2$ are similarly defined. Further we define $I_n^{*}=(I_n+I'_n)/2$,  where $I'_n$ is similarly defined by swapping the role of $\D_1$ and $\D_2$.

Similarly, we can get a consistent estimator $\rho_n$ for $\rho$  by making the sample average of the estimated influence functions zero, i.e.
\begin{align}
	\rho_{n}=\frac{I_{n}}{\sqrt{\hat B_{0}^y\hat B_{0}^z}}.
\end{align}
where
\begin{align*}
	\hat B_{0}^y &= 2n_y^{-1}\sum_{i\in\D_{2y}}
	\hat g'_{1i}(Y_i-\hat m_i)(\hat g_{1i}-\bar{g}_1)+n^{-1}\sum_{i\in\D_2}(\hat g_{1i}-\bar{g}_1)^2,\\
	\hat B_{0}^z &= 2n_z^{-1}\sum_{i\in\D_{2z}}
	\hat g'_{2i}(Z_i-\hat h_i)(\hat g_{2i}-\bar{g}_2) + n^{-1}\sum_{i\in\D_2}(\hat g_{2i}-\bar{g}_2)^2.
\end{align*}
%\begin{align*}
%  & \hat B_{0}^y=\frac{1}{n}\sum_{i\in\D_2}\{2\hat g'_{1i}[Y_i-\hat m(X_i)]+\hat g_{1i}-\bar g_1\}[\hat g_{1i}-\bar g_1];\\
%   &\hat B_{0}^z=\frac{1}{n}\sum_{i\in\D_2}\{2\hat g'_{1i}[Z_i-\hat h(X_i)]+\hat g_{2i}-\bar g_2\}[\hat g_{2i}-\bar g_2].
%\end{align*}
Further $\rho_n^{*}=(\rho_n+\rho'_n)/2$ with $\rho'_n$ being similarly defined by swapping the role of $\D_1$ and $\D_2$.

Before we state the asymptotic behavior of $I_n$ and $\rho_n$, the following regular conditions are required:
\begin{condition}
	\label{condition7}
	$E[\{\hat m_{\D_1}(X)-m(X)\}^{4}]=o(n_y^{-1})$ and $E[\{\hat h_{\D_1}(X)-h(X)\}^{4}]=o(n_z^{-1})$.
\end{condition}
\begin{condition}
	\label{condition8}
	$E[\sigma^2(X)\{\hat m_{\D_1}(X)-m(X)\}^2\{\hat h_{\D_1}(X)-h(X)\}^2]=o(1)$ and $E[\delta^2(X)\{\hat m_{\D_1}(X)-m(X)\}^2\{\hat h_{\D_1}(X)-h(X)\}^2]=o(1)$.
\end{condition}
\begin{condition}
	\label{condition9}
	$E(\sigma^2(X)\{\hat m_{\D_1}(X)-m(X)\}^2[g_2\{h(X)\}-E(g_2)]^2)=o(1)$ and $E(\delta^2(X)\{\hat h_{\D_1}(X)-h(X)\}^2[g_1\{m(X)\}-E(g_1)]^2)=o(1)$.
\end{condition}
Condition \ref{condition7} is stronger than Condition \ref{condition2} because of the link function. Conditions \ref{condition8} and \ref{condition9} are also technical conditions to guarantee the asymptotic normality. When both $\sigma^2(X)$ and $\delta^2(X)$ are finite, Condition \ref{condition8} also holds directly under Condition \ref{condition7}.

\begin{theorem}\label{theorem_asy2}
	Suppose Conditions \ref{condition5}--\ref{condition9}
	are satisfied. We have\\
	{\rm (a)} The asymptotically linear representations  of $I_n^*$ and $\rho^*_n$ respectively are
	\begin{align*}
		I_n^*-I &= \sum_{i\in\D}S^*_{yz,i}+ o_p(N_z^{-1/2}+N_y^{-1/2})\\
		\rho_n^*-\rho &= \sum_{i\in\D}\phi^*_1(O_i)+ o_p(N_z^{-1/2}+N_y^{-1/2}).
	\end{align*}
	Here $$
	S^*_{yz,i}=N_y^{-1}\epsilon_i^*\mathbb{I}(T_{y,i}=1)+N_z^{-1}\eta_i^*\mathbb{I}(T_{z,i}=1)+ N^{-1}\xi^*,$$
	$$
	\phi_1^*(O_i)=\frac{S^*_{yz,i}}{\sqrt{B_0^yB_0^z}}-\rho\frac{S^*_{yy,i}}{2B_0^y}-\rho\frac{S^*_{zz,i}}{2B_0^z},$$
	with $$S^*_{yy,i}=2N_y^{-1}g'_{1i}\{Y_i- m(X_i)\}\{ g_{1i}-E(g_1)\}\mathbb{I}(T_{y,i}=1)+N^{-1}g'_{1i}[\{g_{1i}-E(g_1)\}^2-B_0^y],$$
	$$S^*_{zz,i}=2N_z^{-1}g'_{2i}\{Z_i- h(X_i)\}\{g_{2i}-E(g_2)\}\mathbb{I}(T_{z,i}=1)+N^{-1}g'_{2i}[\{g_{2i}-E(g_2)\}^2-B_0^z].$$
	where $\epsilon_i^*=g'_{1i}\{Y_{i}-m(X_i)\}\{g_{2i}-E(g_2)\},\, \eta_i^*=g'_{2i}\{Z_{i}-h(X_i)\}\{g_{1i}-E(g_1)\}$ and $\xi_i^*=\{g_{1i}-E(g_1)\}\{g_{2i}-E(g_2)\}-I$. And $g_{1i} = g_1\{m(X_i)\},\,g_{2i} = g_2\{h(X_i)\},\,g'_{1i} = g'_1\{m(X_i)\}$ and $g'_{2i} = g'_{2}\{h(X_i)\}$. Recall that $E(g_1) = E[g_1\{m(X)\}]$ and $E(g_2)=E[g_2\{h(X)\}]$.\\
	{\rm (b)}
	The asymptotic normalities of $I_n^*$ and $\rho^*_n$ respectively are
	\begin{align*}
		\sigma^{-1}(I_n^*-I) \overset{d}{\rightarrow} N(0,1) \quad\text{and}\quad
		\sigma_r^{-1}(\rho_n^*-\rho) \overset{d}{\rightarrow} N(0,1).
	\end{align*}
	Here $\sigma^2=\var(\sum_{i\in\D}S^*_{yz,i})$ and $\sigma_r^2=\var\{\sum_{i\in\D}\phi^*_1(O_i)\}$.
\end{theorem}

Based on Theorem \ref{theorem_asy2}, we can similarly derive that the proposed estimators of generalized
genetic covariance and genetic correlation are semiparametric efficient. Obviously, Theorem \ref{theorem_asy2} is also a generalized version of Theorems~\ref{theorem_asy1} and
\ref{theorem_asyp}. Thus, we can similarly conduct further inference procedures, such as the confidence intervals for $I$ and $\rho$, respectively.

\section{Simulation studies}\label{section:sec5}

In this section, we conduct some simulation studies to illustrate our proposed procedures. For estimating $m(X)$ and $h(X)$, we adopt some flexible machine learning algorithms such as least absolute shrinkage and selection operator (LASSO) and the neural network. {For the LASSO method, the estimates are implemented by the \texttt{R} package glmnet while 10-fold cross-validation is used to select corresponding tuning parameters. For the neural network, we apply a multi-layer perceptron neural network (MLP) from \texttt{sklearn}. More specifically, we employ an MLP with two hidden layers, where the number of neurons in each hidden layer is set as 100.} Furthermore, the maximum iteration is 5000 and the learning rate is chosen as  \texttt{adaptive}. In each experiment, we repeat the simulations 500 times.

%To save space, we only consider the response $Y_i,\,Z_{i}$ to be continuous. The simulation results of discrete response is presented in Supplementary Material.
We consider the response $Y_i,\,Z_{i}$ to be continuous or discrete. The continuous response variables are generated from linear models and nonlinear models, respectively. And the discrete response is generated by the logistic regression model. Predictors $X_i$'s are generated from the multivariate normal distribution $\mathcal{N}(0,\Sigma)$ with $\Sigma_{ij}=0.6^{|i-j|}$.

%\subsection{Genetic covariance for continuous response}

%In this subsection, we focus on estimating the genetic covariance and genetic correlation for the~continuous response.
\subsection{Genetic relatedness for continuous response}

	In this subsection, we focus on the finite sample performance of the genetic covariance and genetic correlation estimators with continuous responses.

Firstly, we consider the high-dimensional linear regression models where $p =$ 400, 600 or 800 and $n_y=n_z=400$.
%The response variables $Y_i,\,Z_{i}$ are generated from following linear models,
%\begin{align*}
%	Y_i=X_i^\T\beta+\epsilon_i\quad \text{ and }\quad Z_i=X_i^\T\gamma+\eta_i.
%\end{align*}
%%Given the support $\mathcal{S}=\{j:1\le j \le s\}$,
%The signals of $\beta$ satisfy that $\beta_j=0.4(1+j/2s_1)$ for $1\leq j\leq s_1$, and signals of $\gamma$ satisfy that $\gamma_j=0.3(1-j/2s_2)$ for $1\leq j\leq s_2$.
%The rest elements of $\beta$ and $\gamma$ are zero.
%Here $s_1$ and $s_2$ are the sparsity levels of $\beta$ and $\gamma$, respectively. Besides, error terms $\epsilon_i$ and $\eta_i$ are generated from the standard normal distribution. In this model setting, the genetic covariance $I=\beta^\T\Sigma\gamma.$
\begin{example}
	\label{exp1}
	The continuous outcomes are generated from following linear model,
	\begin{align*}
		Y_i=X_i^\T\beta+\epsilon_i, \text{ and } Z_i=X_i^\T\gamma+\eta_i.
	\end{align*}
	%Given the support $\mathcal{S}=\{j:1\le j \le s\}$,
	The signals of $\beta$ satisfy that $\beta_j=0.4(1+j/2s_1)$, and signals of $\gamma$ satisfy that $\gamma_j=0.3(1-j/2s_2)$.
	Here $s_1$ and $s_2$ are the sparsity levels of $\beta$ and $\gamma$, respectively. Besides, error terms $\epsilon_i$ and $\eta_i$ are generated from the standard normal distribution $N(0,1)$. In this example, the genetic covariance $I=\beta^\T\Sigma\gamma.$
\end{example}

We investigate the consistency of $\tilde I_N$ at first. The results are presented in Table \ref{tab1} with overlapping setting, i.e. the data of two traits are from the same group of individuals, while Table~\ref{tab2} with non-overlapping setting
, i.e. the data of two traits are from different groups of individuals with $N_0=0$. Each row reports absolute value of average bias and relative bias with different sparsity levels and dimension size settings. Here, we define relative bias (denoting as $\mbox{rBIAS}$) as absolute value of average bias divides the true value $I$. We compare two approaches which are applied to estimate $E(Y\mid X)$ and $E(Z\mid X)$: the MLP and the LASSO method. From table \ref{tab1}, we find that our method works well and the bias is relatively small no matter how $p, s_1$ and $s_2$ change. From Table 2, we find that the relative bias is acceptable no matter how $p, s_1$ and $s_2$ change. Besides, the estimated results by MLP method can be as good as LASSO method in both overlapping and non-overlapping cases. From the results, it is clear that $\tilde I_N$ is consistent with small relative bias.

\begin{table}
\footnotesize
\renewcommand\arraystretch{0.8}
\centering \tabcolsep 4pt \LTcapwidth 6in
	\def~{\hphantom{0}}
	\caption{Performance of estimated genetic covariance $\tilde I_N$ when $m(X)$ is consistently estimated and $h(X)$ is estimated as 0: \textbf{Example \ref{exp1} with overlapping setting}}{%
\scalebox{1}{		
\begin{tabular}{*{10}{c}}
\toprule {}&{}&{}&{}&\multicolumn{2}{c}{MLP}&\multicolumn{2}{c}{LASSO}\\
\midrule
			{$n_y$}& {$n_z$}& {$p$}& {($s_1,s_2$)}  & $\mbox{BIAS}$ & $\mbox{rBIAS}$ & $\mbox{BIAS}$ & $\mbox{rBIAS}$  \\[5pt]
\midrule
			400 &  400  &   400   & (20,20) & 0.04201 & 0.0053 & 0.1305 & 0.0164\\
			&      &      & (30,30) & 0.0570 & 0.0046
			& 0.0460 & 0.0037 \\
			&      &      & (40,40) & 0.0643  & 0.0038
			& 0.1456 & 0.0087 \\
\midrule
			400 & 400 & 600 & (20,20)& 0.0218 & 0.0027
			& 0.0279 & 0.0035 \\
			&      &      & (30,30) & 0.0334 & 0.0027
			& 0.1698 & 0.0137 \\
			&      &      & (40,40) & 0.0757 & 0.0045
			& 0.1236 & 0.0074 \\
\midrule
		400 & 400 & 800 & (20,20) & 0.0091 & 0.0012
		& 0.0941 & 0.0118 \\
		&      &      & (30,30) & 0.0628 & 0.0051
		& 0.1072 & 0.0087 \\
		&      &      & (40,40) & 0.0237 & 0.0014
		& 0.2273 &  0.0136\\
\bottomrule
	\end{tabular}}
	\label{tab1}
\footnotesize
	  \begin{tablenotes}
	\item MLP: estimated by MLP method; LASSO: estimated by Lasso method.
	 \end{tablenotes}
}
\end{table}

\begin{table}
\footnotesize
\renewcommand\arraystretch{0.8}
\centering \tabcolsep 4pt \LTcapwidth 6in
 	\def~{\hphantom{0}}
	\caption{Performance of estimated genetic covariance $\tilde I_N$ when $m(X)$ is consistently estimated and $h(X)$ is estimated as 0:  \textbf{Example \ref{exp1} with non-overlapping setting}}{%
\scalebox{1}{		
\begin{tabular}{*{10}{c}}
\toprule {}&{}&{}&{}&\multicolumn{2}{c}{MLP}&\multicolumn{2}{c}{LASSO}\\
\midrule
			{$n_y$}& {$n_z$}& {$p$}& {($s_1,s_2$)}  & $\mbox{BIAS}$ & $\mbox{rBIAS}$ & $\mbox{BIAS}$ & $\mbox{rBIAS}$ \\[5pt]
  \midrule
      400 & 400 & 400 & (20,20) & 0.1317 & 0.0166
      & 0.1465 & 0.0184   \\
      &      &      & (30,30) & 0.1114  & 0.0090
      & 0.2661 &  0.0215   \\
      &      &      & (40,40) & 0.1891 & 0.0113
      & 0.3790 &  0.0226   \\
  \midrule
      400 & 400 & 600 & (20,20) & 0.1003 & 0.0126
      & 0.1416 &  0.0178   \\
      &      &      & (30,30) & 0.1296 & 0.0105
      & 0.2456 &  0.0199   \\
      &      &      & (40,40) & 0.0179 & 0.0011
      & 0.3697 & 0.0221    \\
      \midrule
      400 & 400 & 800 & (20,20) & 0.1123 & 0.0141
      & 0.1838 &  0.0231   \\
      &      &      & (30,30) & 0.2033 & 0.0165
      & 0.2993 &  0.0242   \\
      &      &      & (40,40) & 0.1160  & 0.0069
      & 0.4233 & 0.0253  \\
  \bottomrule
	\end{tabular}}
	\label{tab2}
\footnotesize
	  \begin{tablenotes}
	\item MLP: estimated by MLP method; LASSO: estimated by Lasso method.
	 \end{tablenotes}
}
\end{table}

Next we investigate the performance of $I_n^*$. In addition, we also compare our method with the approach proposed by \cite{wang2021unified}. The results of our proposed methods are summarized in Table \ref{tab3} and Table ~\ref{tab4}. Each row reports empirical coverage probability (denoting as $\mbox{CP}$), absolute value of average bias (denoting as $\mbox{BIAS}$) and estimated confidence interval length (denoting as $\mbox{LEN}$) with different sparsity levels and dimension size settings. In this part, we set $s_1=s_2=s$.
From Tables \ref{tab3}-\ref{tab4}, we have the following findings. Firstly, we can see that our method performs well for different values of $p$ and $s$. The coverage probability is always around 0.95 and the bias is relatively small no matter how $p$ and $s$ change. Secondly, for the high-dimensional linear regression models, both LASSO and MLP methods can exhibit good performance.
%under both overlapping and non-overlapping cases. MLP method performs as well as Lasso method in the overlapping case, while Lasso method can get better coverage probability, smaller bias and shorter confidence interval than MLP method when the data setting is non-overlapping.
Thirdly, compared with the method in \cite{wang2021unified}, our method can reach a smaller bias and shorter confidence interval when both coverage probabilities are approximately 0.95.
%In the non-overlapping case, we see that the coverage probability for the method proposed by \cite{wang2021unified} is smaller than 0.90 as $p$ and $s$ increase, and its confidence interval lengths are larger than that of ours.

\begin{table}
\footnotesize
\renewcommand\arraystretch{0.8}
\centering \tabcolsep 4pt \LTcapwidth 6in
	\def~{\hphantom{0}}
		\caption{Performance of estimated genetic covariance $ I_n^*$  generated from \textbf{Example \ref{exp1} with overlapping setting} }{%
\scalebox{1}{
		\begin{tabular}{*{12}{c}}
		\toprule {}&{}&{}&\multicolumn{3}{c}{MLP}&\multicolumn{3}{c}{LASSO}
			&\multicolumn{3}{c}{Wang}\\
\midrule
			{$n$}& {$p$}& {$s$}  & $\mbox{CP}$ & $\mbox{BIAS}$ & $\mbox{LEN}$ & $\mbox{CP}$ & $\mbox{BIAS}$ & $\mbox{LEN}$ & $\mbox{CP}$ & $\mbox{BIAS}$ & $\mbox{LEN}$\\[5pt]
\midrule
			400 & 400 & 10 & 0.930 & 0.027 & 0.870
			& 0.924 & 0.017 & 0.851
			& 0.944 & 0.044 & 1.232 \\
			&       & 20 & 0.946 & 0.003 & 1.733
			& 0.944 & 0.004  & 1.729
			& 0.946 & 0.097 & 2.513 \\
			&       & 30 & 0.950 & 0.081 & 2.587
			& 0.958 & 0.015  & 2.595
			& 0.950 & 0.026  & 3.721 \\
\midrule
			400 &  600  & 10 & 0.944 & 0.011  & 0.913
			& 0.934 & 0.006  & 0.851
			& 0.928 & 0.065  & 1.241 \\
			&       & 20 & 0.956 & 0.073  & 1.721
			& 0.950 & 0.030  & 1.717
			& 0.950 & 0.009  & 2.480 \\
			&       & 30 & 0.932 & 0.102 & 2.066
			& 0.950 & 0.003 & 2.615
			& 0.958 & 0.093 & 3.718 \\
\midrule
			400 & 800 & 10 & 0.936 & 0.046  & 0.866
			& 0.940 & 0.023  & 0.855
			& 0.930 & 0.047  & 1.228 \\
			&       & 20 & 0.948 & 0.061 & 1.725
			& 0.952 & 0.032 & 1.717
			& 0.942 & 0.113 & 2.499 \\
			&       & 30 & 0.944 & 0.096 & 2.587
			& 0.940 & 0.091 & 2.575
			& 0.938 & 0.144 & 3.728\\
\bottomrule
	\end{tabular}}

	\label{tab3}
\footnotesize
	  \begin{tablenotes}
	\item  MLP: estimated by MLP method; LASSO: estimated by Lasso method; Wang: the method proposed by \cite{wang2021unified}.
	 \end{tablenotes}
}
\end{table}
	
%	We next explore the simulations of estimating the genetic correlation for the continuous response with linear models.

\begin{table}
\footnotesize
\renewcommand\arraystretch{0.8}
\centering \tabcolsep 4pt \LTcapwidth 6in
	\def~{\hphantom{0}}
		\caption{Performance of estimated genetic covariance $ I_n^*$ generated from \textbf{Example \ref{exp1} with non-overlapping setting} }{%
\scalebox{1}{
		\begin{tabular}{*{12}{c}}
		\toprule {}&{}&{}&\multicolumn{3}{c}{MLP}&\multicolumn{3}{c}{LASSO}
			&\multicolumn{3}{c}{Wang}\\
\midrule
			{$n$}& {$p$}& {$s$}  & $\mbox{CP}$ & $\mbox{BIAS}$ & $\mbox{LEN}$ & $\mbox{CP}$ & $\mbox{BIAS}$ & $\mbox{LEN}$ & $\mbox{CP}$ & $\mbox{BIAS}$ & $\mbox{LEN}$\\[5pt]
  \midrule
    400 &   400   & 10 & 0.922 & 0.025 & 0.808
                  & 0.922 & 0.011 & 0.741
                  & 0.926 & 0.049 & 0.904\\
       &     & 20 & 0.930 & 0.038 & 1.443
                  & 0.940 & 0.003 & 1.376
                  & 0.888 & 0.127 & 1.772\\
      &      & 30 & 0.924 & 0.044 & 2.070
                  & 0.954 & 0.028 & 1.999
                  & 0.920 & 0.201 & 2.638 \\
  \midrule
    400 & 600 & 10 & 0.924 & 0.016  & 0.815
                   & 0.938 & 0.013  & 0.753
                   & 0.930 & 0.059  & 0.895 \\
       &      & 20 & 0.944 & 0.035  & 1.439
                   & 0.942 & 0.022  & 1.380
                   & 0.930 & 0.119  & 1.777\\
       &      & 30 & 0.932 & 0.030  & 2.066
                   & 0.940 & 0.017  & 2.011
                   & 0.900 & 0.276  & 2.628\\
      \midrule
     400 & 800 & 10 & 0.938 & 0.023  & 0.815
                   & 0.922 & 0.016  & 0.745
                   & 0.908 & 0.077  & 0.893 \\
       &      & 20 & 0.926 & 0.035  & 1.439
                   & 0.946 & 0.024  & 1.380
                   & 0.888 & 0.206  & 1.759 \\
       &      & 30 & 0.918 & 0.033  & 2.066
                   & 0.944 & 0.060  & 2.003
                   & 0.892 & 0.287  & 2.618 \\
  \bottomrule
	\end{tabular}}

	\label{tab4}
\footnotesize
	  \begin{tablenotes}
	\item  MLP: estimated by MLP method; LASSO: estimated by Lasso method; Wang: the method proposed by \cite{wang2021unified}.
	 \end{tablenotes}
}
\end{table}

We next explore the simulations of estimating the genetic correlation for the continuous response with linear models.
	
	\begin{example}
		\label{exp2}
		The continuous responses are generated from following linear models.
		$$Y_i=X_i^\T\beta+\epsilon_i\ \mbox{and} ~  Z_i=X_i^\T\gamma+\eta_i.$$
		Here, the coefficients of $\beta$ are generated from $\beta_i=1+i/2s_{1}$ for $ 1 \leq i \leq s_{1}$, and the signals of $\gamma$ satisfy that $\gamma_j=2(1-j/2s_{2})$ for $ s_{1}+1 \leq j \leq s_{1}+s_{2}$, and $\beta_j=0$, $\gamma_j=0$ otherwise. $s_{1}$ and $s_{2}$ are the sparsity levels of $\beta$ and $\gamma$, respectively. The error terms $\epsilon_i$ and $\eta_i$ are generated from the normal distribution $N(0,1)$ as before. In this example, the genetic correlation is $\rho=\beta^\T\Sigma\gamma/\sqrt{\beta^\T\Sigma\beta\cdot \gamma^\T\Sigma\gamma}.$
	\end{example}
	
	We set the sample size $n_y=n_z=400$, and the dimension $p = 400, 600$ or 800. We consider different sparsity parameters $s_{1}=20, 25, 30$ and $s_{2}=$ 5. The estimated results of genetic correlation $\rho$ with overlapping setting and non-overlapping setting are presented in Table \ref{tab5} and Table \ref{tab6}, respectively. Firstly, we see that our proposed method shows some pretty good results and coverage probabilities are approximately 0.95 and absolute biases are relatively small when $p$ and $s_1$ are considered for different settings. Secondly, we can know that the coverage probability for MLP method approaches to 0.95 and the absolute bias is under control, which is similar to LASSO method. Therefore, we conclude that MLP method can provide accurate estimation results despite its absolute bias and confidence interval lengths are larger than that of LASSO method. Thirdly, in general, we notice that the biases and the confidence interval lengths are smaller with non-overlapping setting than those with overlapping setting. The reason is that the total sample size of non-overlapping is larger than that of overlapping setting although they have the same sample setting $n_y=n_z=400$.
	%So the estimated results of the latter one are more approached to those of the oracle method.
	
	\begin{table}[!h]
\footnotesize
\renewcommand\arraystretch{0.8}
\centering \tabcolsep 4pt \LTcapwidth 6in
		\def~{\hphantom{0}}
		\caption{Performance of estimated genetic correlation $ \rho_n^*$
%			by our proposed method with continuous outcomes
			generated from \textbf{Example \ref{exp2} with overlapping setting}}{%
\scalebox{1}{			
\begin{tabular}{*{9}{c}}
				%\\
				\toprule				{}&{}&{}&\multicolumn{3}{c}{MLP}&\multicolumn{3}{c}{LASSO} \\
\midrule
				{$n$}& {$p$}& {$(s_{1},s_{2})$}  & $\mbox{CP}$ & $\mbox{BIAS}$ & $\mbox{SE}$  & $\mbox{CP}$ & $\mbox{BIAS}$ & $\mbox{SE}$ \\[5pt]
\midrule
				400 & 400 & (20,5) & 0.926 & 0.0161 & 0.0408
				& 0.928 & 0.0003 & 0.0363  \\
				&      & (25,5) & 0.938 & 0.0160 & 0.0405
				& 0.944 & 0.0006 & 0.0363 \\
				&      & (30,5) & 0.944 & 0.0121 & 0.0406
				& 0.950 & 0.0005 & 0.0365 \\
\midrule
				400 & 600 & (20,5) & 0.926 & 0.0155 & 0.0409
				& 0.942 & 0.0002 & 0.0363\\
				&       & (25,5) & 0.952 & 0.0105 & 0.0406
				& 0.940 & 0.0014 & 0.0364 \\
				&      & (30,5) & 0.926 & 0.0127 & 0.0406
				& 0.950  & 0.0004 & 0.0364 \\
\midrule
				400 & 800 & (20,5) & 0.938 & 0.0155 & 0.0409
				& 0.950 & 0.0029 & 0.0363   \\
				&       & (25,5) & 0.924 & 0.0157 & 0.0404
				& 0.936 & 0.0002 & 0.0364 \\
				&      & (30,5) & 0.932 & 0.0133 & 0.0412
				& 0.926 & 0.0004 & 0.0365\\
\bottomrule
		\end{tabular}}
		\label{tab5}
\footnotesize
		\begin{tablenotes}
		\item	MLP: estimated by the MLP method; LASSO: estimated by the Lasso method.
		\end{tablenotes}
}
	\end{table}
	
	\begin{table}[!h]
\footnotesize
\renewcommand\arraystretch{0.8}
\centering \tabcolsep 4pt \LTcapwidth 6in
		\def~{\hphantom{0}}
		\caption{Performance of estimated genetic correlation $ \rho_n^*$
%			by our proposed method with continuous outcomes
			generated from \textbf{Example \ref{exp2} with non-overlapping setting}}{%
\scalebox{1}{
			\begin{tabular}{*{12}{c}}
				%\\
				\toprule				{}&{}&{}&\multicolumn{3}{c}{MLP}&\multicolumn{3}{c}{LASSO} \\
\midrule
				{$n$}& {$p$}& {$(s_{1},s_{2})$}  & $\mbox{CP}$ & $\mbox{BIAS}$ & $\mbox{SE}$  & $\mbox{CP}$ & $\mbox{BIAS}$ & $\mbox{SE}$  \\[5pt]
\midrule
				400& 400 & (20,5) & 0.922  & 0.0088 & 0.0329
				& 0.934 & 0.0010 & 0.0272 \\
				&      & (25,5) & 0.926 & 0.0079 & 0.0330
				& 0.932 & 0.0010 &	0.0271 \\
				&      & (30,5) & 0.922 & 0.0050 & 0.0330
				& 0.940 & 0.0015 & 0.0270 \\
\midrule
				400 & 600 & (20,5) & 0.924 & 0.0073 & 0.0331
				& 0.928	& 0.0012 & 0.0272 \\
				&      & (25,5) & 0.942 & 0.0067 & 0.0329
				& 0.966 & 0.0009 & 0.0271 \\
				&      & (30,5) & 0.954 & 0.0047 & 0.0326
				& 0.950	& 0.0020 & 0.0270  \\
\midrule
				400 & 800 & (20,5) & 0.924 & 0.0094 & 0.0331
				& 0.930 & 0.0016 & 0.0272  \\
				&      & (25,5) & 0.940 & 0.0069 & 0.0331
				& 0.938	& 0.0017 & 0.0271 \\
				&      & (30,5) & 0.932 & 0.0044 & 0.0327
				& 0.942 & 0.0023 & 0.0271\\
\bottomrule
		\end{tabular}}
		\label{tab6}
\footnotesize
		\begin{tablenotes}
	\item		 MLP: estimated by the MLP method; LASSO: estimated by the Lasso method.
		\end{tablenotes}
}
	\end{table}

	Next, we present the simulation results for genetic covariance with nonlinear model setting.
%The nonlinear functions are taken from \cite{wang2021unified}.
The error terms $\epsilon_i$ and $\eta_i$ are generated from the standard normal distribution.

 	\begin{example}
		\label{exp3}
	The continuous outcome $Y_i$ is generated from a nonlinear model
	\begin{align*}
		Y_i=-5+2\sin(\pi x_{i1}x_{i2})+4(x_{i3}-0.5)^2+2x_{i5}+x_{i6}+\epsilon_i,
	\end{align*}
	%For the continuous outcome $z_i$, we consider two different generating functions. One is linear model and another is nonlinear model.
	and $Z_i$ is generated from another nonlinear model
	\begin{align*}
		Z_i=3\sin(x_{i1})+x_{i3}^3+\exp(x_{i4})+\eta_i.
	\end{align*}
	Here $x_j, j=1,\cdots,6$ are elements of $X$. Recall that predictors $X_i$'s are generated from the multivariate normal distribution $\mathcal{N}(0,\Sigma)$ with $\Sigma_{ij}=0.6^{|i-j|}$.
\end{example}%
%	\begin{example}
%		\label{exp3}
%		The continuous outcome $Y_i$ is generated from a nonlinear model
%		\begin{align*}
%			Y_i=-5+2\sin(\pi x_{i1}x_{i2})+4(x_{i3}-0.5)^2+2x_{i5}+x_{i6}+\epsilon_i,
%		\end{align*}
%		%For the continuous outcome $z_i$, we consider two different generating functions. One is linear model and another is nonlinear model.
%		$Z_i$ is generated from another nonlinear model
%		\begin{align*}
%			Z_i=3\sin(x_{i1})+x_{i3}^3+\exp(x_{i4})+\eta_i.
%		\end{align*}
%		Here $x_j, j=1,\cdots,6$ are elements of $X$. Recall that predictors $X_i$'s are generated from the multivariate normal distribution $\mathcal{N}(0,\Sigma)$ with $\Sigma_{ij}=0.6^{|i-j|}$.
%	\end{example}
%	
	Table \ref{tab7} reports empirical results based on nonlinear models.
 The columns indexed by Overlap and Non-overlap denote that the two data sets are from the same samples and from different samples, respectively.
	Each row reports empirical coverage probability (CP), absolute value of average bias (BIAS) and estimated standard deviation (SE) with different dimension size settings. We apply MLP method and LASSO method to estimate nonlinear models. In addition, we also compare our method with the method proposed by \cite{wang2021unified}. We observe that the coverage probability of our proposed method with MLP method can still approximately achieve the desirable 0.95 level and it is robust for different $p$'s. However, our results with LASSO method suffer from biased estimations and the estimated coverage probability is under 0.90 and the absolute bias is much larger than that of MLP method. As a comparison, we know that the proposed method by \cite{wang2021unified} with LASSO method performs worse than our method with MLP algorithm with much larger bias and smaller coverage probabilities for some cases. Therefore, under nonlinear model setting, MLP method performs better than LASSO method in both overlapping and non-overlapping cases.

	\begin{table}
\footnotesize
\renewcommand\arraystretch{0.8}
\centering \tabcolsep 4pt \LTcapwidth 6in
		\def~{\hphantom{0}}
			\caption{Performance of estimated genetic covariance $ I_n^*$ generated from \textbf{Example \ref{exp3}}}{%
\scalebox{1}{
			\begin{tabular}{*{12}{c}}
\toprule				{}&{}&{}&\multicolumn{3}{c}{MLP}&\multicolumn{3}{c}{LASSO}
				& \multicolumn{3}{c}{Wang}\\
\midrule
				{}&{$n$}& {$p$}& $\mbox{CP}$ & $\mbox{BIAS}$ & $\mbox{SE}$ & $\mbox{CP}$ & $\mbox{BIAS}$ & $\mbox{SE}$ & $\mbox{CP}$ & $\mbox{BIAS}$ & $\mbox{SE}$ \\[5pt]
\midrule
				Overlap & 400 & 400 & 0.926 & 0.555 & 3.206	
				& 0.846 & 2.727 & 2.435
				& 0.938 & 1.995 & 3.522 \\
				&  & 600 & 0.928 & 0.513 & 3.258
				& 0.866 & 2.892 & 2.460
				& 0.930 & 2.338 & 3.667 \\
				& & 800 & 0.928 & 0.558 & 3.206
				& 0.858 & 2.837 & 2.434
				& 0.918 & 2.062 & 3.506 \\
				&     & 1000 & 0.932 & 0.676 & 3.220	
				& 0.862 & 2.833 & 2.438
				& 0.930 & 2.212 & 3.645 \\
\midrule
				Non-overlap & 400 & 400 & 0.918	& 0.230 & 2.713
				& 0.884 & 2.441 & 2.467
				& 0.872 & 2.552 & 3.158 \\
				& & 600 & 0.910 & 0.605 & 2.781
				& 0.854 & 2.475 & 2.450
				& 0.898 & 2.274 & 3.119 \\
				&     & 800 & 0.912	& 0.519 & 2.766
				& 0.862 & 2.360 & 2.445
				& 0.924 & 2.336 & 3.172 \\
				&     & 1000 & 0.912 & 0.478 & 2.710
				& 0.882 & 2.123 & 2.046
				& 0.870 & 2.401 & 3.080\\
\bottomrule
		\end{tabular}}
		\label{tab7}
\footnotesize
		\begin{tablenotes}
 \item			Overlap: two data sets are from the same samples; Non-overlap: two data sets are from different samples. \\
    MLP: estimated by MLP method; LASSO: estimated by Lasso method; Wang: the method proposed by \cite{wang2021unified}.
		\end{tablenotes}
}
	\end{table}

	%Overall, numerical results show the superiority of our proposed methods. We notice that MLP method performs as well as LASSO method under linear models
%	and logistic models 	but MLP can achieve more accurate estimated results than LASSO under nonlinear models.
%	Moreover, comparing non-overlapping setting with overlapping setting, the coverage probabilities are lower due to shorter confidence intervals.

	\subsection{Genetic covariance for discrete response}
	In this subsection, we consider the finite sample performance of proposed estimators with discrete responses.
	
	\begin{example}
		\label{exp4}
		The binary outcomes $Y_i$ and $Z_i$ are generated by the following logistic regression models,
		\begin{align*}
			\Pr(Y_i=1|X_i)=\frac{\exp(X_i^\T\beta)}{1+\exp(X^\T_i\beta)},\,\,\Pr(Z_i=1|X_i)=\frac{\exp(X^\T_i\gamma)}{1+\exp(X^\T_i\gamma)}.
		\end{align*}
		We consider the high-dimensional logistic regression model where $p = 400, 600$ or 800 and $n_y=n_z=400$. The coefficient vectors $\beta$ and $\gamma$ are similar to linear models. For the true regression coefficients, given the support $\mathcal{S}=\{j:1\le j \le s\}$, the signals of $\beta$ satisfy that $\beta_j=0.2(1+j/2s)$, and the signals of $\gamma$ follow that $\gamma_j=0.3(1-j/2s)$, for all $j \in \mathcal{S}$.
	\end{example}
	The numerical results are summarized in Table \ref{tab8} with overlapping case and Table \ref{tab9} with non-overlapping case. Each row reports empirical coverage probability (CP), absolute value of average bias (BIAS) and estimated confidence interval length (LEN) with different dimension size settings and different sparsity levels of $\beta$ and $\gamma$ as $(s_1, s_2) \in \{(5,5),(10,10)\}$. We compare two approaches applied to estimate functions: MLP algorithm and Lasso method. In addition, we also compare our method with the approach proposed by \cite{ma2022statistical}.
	
	We have the following observations. Firstly, all estimated results are under control of our method when $p$ and $s$ change. The coverage probability is always around the normal level and the bias is relatively small. Secondly, there is no significant difference between the performance of MLP method and Lasso method. Specially, in the overlapping case, Lasso method can get better coverage probability, smaller bias and shorter confidence interval length than MLP method. For non-overlapping setting, MLP method leads to smaller bias and shorter confidence interval length than Lasso method when the coverage probabilities reach a similar level. Thirdly, our method achieves more satisfying performance than the estimated method proposed by \cite{ma2022statistical} in all settings. For both overlapping case and non-overlapping case, the coverage probabilities of our method are approximately 0.95 while their results are around 0.90. In addition, the bias is much smaller and the confidence interval width is much shorter in our method than those in theirs.
	%Therefore, our method performs pretty well in logistic models.

	\begin{table}[!h]
\footnotesize
\renewcommand\arraystretch{0.8}
\centering \tabcolsep 4pt \LTcapwidth 6in
		\def~{\hphantom{0}}
		\caption{Performance of estimated genetic covariance $ I_n^*$
%			by our proposed method with binary outcomes
			generated from \textbf{Example \ref{exp4} with overlapping setting}}{%
 \scalebox{1}{
			\begin{tabular}{*{12}{c}}

\toprule				%\\
{}&{}&{}&\multicolumn{3}{c}{MLP}&\multicolumn{3}{c}{LASSO}
				&\multicolumn{3}{c}{Ma}\\
\midrule
				{$n$}& {$p$}& {$s$}  & CP & BIAS & LEN & CP & BIAS & LEN & CP & BIAS & LEN \\[5pt]
\midrule
				400 & 400 & 5 & 0.930 & 0.023 & 0.463
				& 0.934 & 0.018 & 0.349
				& 0.782 & 0.598 & 1.685 \\
				&  & 10 & 0.934 & 0.065 & 1.125
				& 0.944 & 0.076 & 0.886
				& 0.800 & 1.171 & 2.991 \\
				% &      & 15 & 0.934 & 0.132 & 1.788
				%            & 0.928 & 0.178 & 1.725 \\
\midrule
				400 & 600 & 5 & 0.942 & 0.022  & 0.447
				& 0.952 & 0.015  & 0.345
				& 0.800 & 0.769  & 2.058\\
				&  & 10 & 0.944 & 0.071  & 1.121
				& 0.956 & 0.080  & 0.878
				& 0.744 & 1.403  & 3.397\\
				%   &      & 15 & 0.954 & 0.150  & 1.858
				%               & 0.948 & 0.220  & 1.756 \\
\midrule
				400 & 800 & 5 & 0.938 & 0.020  & 0.482
				& 0.954 & 0.009  & 0.349
				& 0.778 & 0.858  & 2.262 \\
				& & 10 & 0.940 & 0.075  & 0.933
				& 0.962 & 0.064  & 0.882
				& 0.728 & 1.514  & 3.554\\
\bottomrule
		\end{tabular}}
		\label{tab8}
\footnotesize
		\begin{tablenotes}
		 \item	MLP: estimated by MLP method; LASSO: estimated by Lasso method; Ma: the method proposed by \cite{ma2022statistical}.
		\end{tablenotes}
}
	\end{table}
	
	\begin{table}[!h]
\footnotesize
\renewcommand\arraystretch{0.8}
\centering \tabcolsep 4pt \LTcapwidth 6in
		\def~{\hphantom{0}}
		\caption{Performance of estimated genetic covariance $ I_n^*$
%			by our proposed method with binary outcomes
			generated from \textbf{Example \ref{exp4} with non-overlapping setting}}{%
\scalebox{1}{
			\begin{tabular}{*{12}{c}}
				%\\				\\
				\toprule {}&{}&{}&\multicolumn{3}{c}{MLP}&\multicolumn{3}{c}{LASSO}
				&\multicolumn{3}{c}{Ma}\\
\midrule
				{$n$}& {$p$}& {$s$}  & CP & BIAS & LEN & CP & BIAS & LEN & CP & BIAS & LEN\\[5pt]
\midrule
				400 & 400 & 5 & 0.938 & 0.017 & 0.474
				& 0.942 & 0.028 & 0.517
				& 0.860 & 0.535 & 1.671 \\
				&  & 10 & 0.932 & 0.089 & 1.094
				& 0.924 & 0.139 & 1.384
				& 0.854 & 1.091 & 3.067 \\
				%  &      & 15 & 0.922 & 0.118 & 1.764
				%              & 0.942 & 0.510 & 3.038  \\
\midrule
				400 & 600 & 5 & 0.934 & 0.025 & 0.478
				& 0.942 & 0.029 & 0.521
				& 0.832 & 0.660 & 2.040 \\
				& & 10 & 0.946 & 0.076 & 1.129
				& 0.928 & 0.149 & 1.392
				& 0.874 & 1.216 & 3.402 \\
\midrule
				%   &      & 15 & 0.910 & 0.128 & 1.744
				%               & 0.944 & 0.523 & 3.062  \\
				400 & 800 & 5 & 0.926 & 0.025 & 0.470
				& 0.944 & 0.035 & 0.521
				& 0.888 & 0.707 & 2.264 \\
				&  & 10 & 0.912 & 0.053 & 1.070
				& 0.928 & 0.162 & 1.407
				& 0.890 & 1.238 & 3.592\\
\bottomrule
		\end{tabular}}
		\label{tab9}
\footnotesize
		\begin{tablenotes}
		\item	MLP: estimated by MLP method; LASSO: estimated by Lasso method; Ma: the method proposed by \cite{ma2022statistical}.
		\end{tablenotes}
}
	\end{table}
	
Overall, numerical results show the superiority of our proposed methods. We notice that MLP method performs as well as Lasso method under linear models and logistic models but MLP can achieve more accurate estimated results than Lasso under nonlinear models.
%Moreover, comparing non-overlapping setting with overlapping setting, the coverage probabilities are lower due to shorter confidence intervals.

	\section{Real data analysis}\label{section:sec6}
	
	To demonstrate the usefulness of our methods, we analyze a Carworth Farms (CFW) White mice data set reported in \cite{Parker2016}. According to \cite{Parker2016}, 1,200 male CFW mice were used to perform a genome-wide association study (GWAS) of behavioral, physiological and gene expression phenotypes. These CFW mice were phenotyped for conditioned fear, anxiety behavior, methamphetamine sensitivity, prepulse inhibition, fasting glucose levels, body weight, tail length, testis weight and so on. The phenotypes can be classified into three categories: behavioral, physiological and expression quantitative traits. Besides, we can regard these mice as independent individuals since the CFW mice do not have cryptic relatedness.
	
	\cite{Parker2016} carried out a series of experiments to measure behavioral traits. For methamphetamine sensitivity traits, the locomotor activity of each mouse was measured by time spent in the center of the arena (in seconds) after the injection in each day. The time spent in the center of the arena on the first day provides a measure of baseline response to a novel environment. The time recorded on the third day provides a measure of methamphetamine sensitivity. The conditioned fear traits measure the recall of the fearful memory by measuring the mice' freezing behavior in response to the stimulus. The physiological traits include body weights taken during methamphetamine sensitivity testing and the weights of various muscles.
	
	There were health concerns or other concerns with some mice, and we exclude these observations from further analysis. The observation flagged as a possible sample mixup due to mishandling or mislabeling of the flowcells is also removed. The data set includes various levels of missingness in the phenotypes, so we choose 24 phenotypes with the fewest missingness. After the pre-processing, the data set consists of 898 mice with 92734 genetic variants (SNPs) and 24 different phenotypes. We have calculated the pair-wise genetic correlation for 276 pairs of the traits and reported in Figure 1.
	\begin{figure}[t]
\begin{center}
\includegraphics[width=5in]{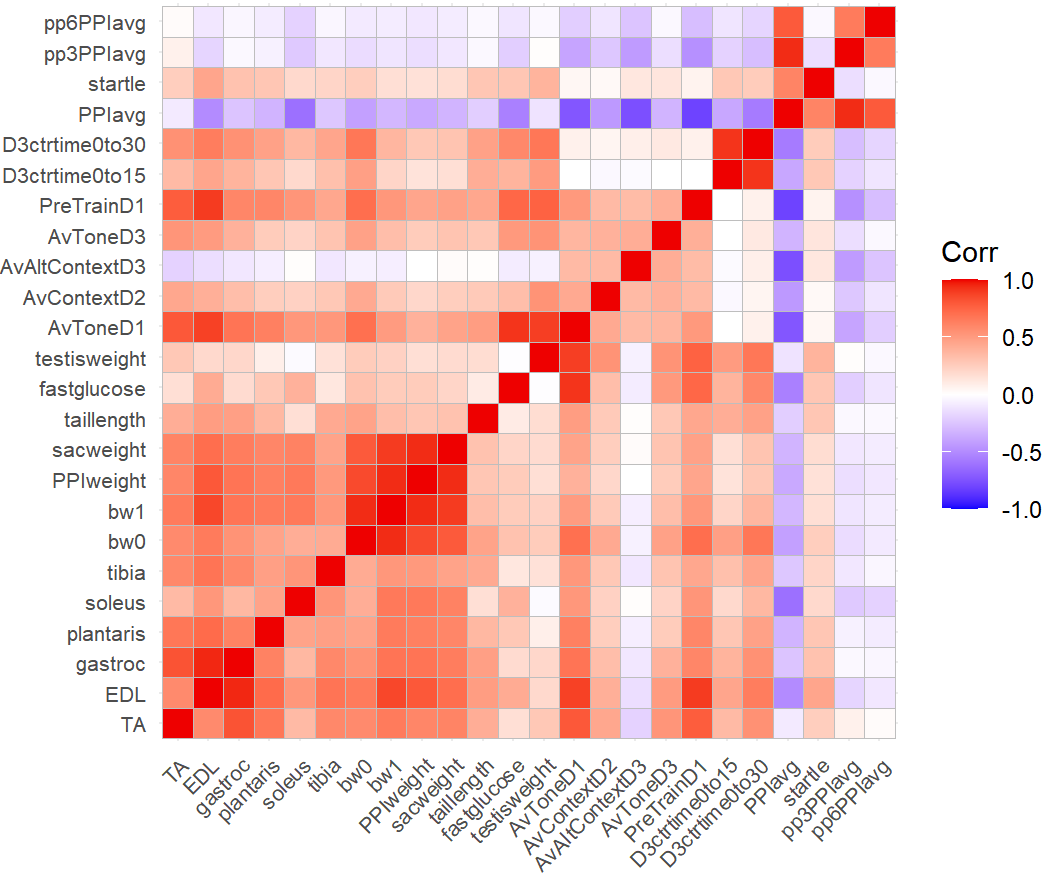}
\end{center}
	\caption{Heatmap of phenotypic correlation for 24 phenotypes.}
		\label{fig1}
\end{figure}

	%\begin{figure}[t]
%		% The arguments in the next line are {height}{optional width}{used only by OUP for typesetting} for figure empty box eg
%		%\figurebox{20pc}{25pc}{}
%		%if actual size of graphics need plese see below command
%		%\figurebox{}{}{}[fig1]
%		%need to reducing the figure size use below command
%		%\figuresize{.8}%
%		\figuresize{.6}
%		\figurebox{20pc}{25pc}{}[heatmap_1206.jpg]
%		% note that files may not be rotated
%		\caption{Heatmap of phenotypic correlation for 24 phenotypes.}
%		\label{fig1}
%	\end{figure}
	
	%{\color{red}Estimation of $\hat \beta$ and $\hat \gamma$ are implemented by package glmnet with 10-fold cross-validation. }
	
	We can obtain several important conclusions from our estimated results for the real data. Firstly, the heatmap indicates that these phenotypic correlation coefficients are usually positive since the most blocks in the heatmap are red. Secondly, physiological traits have a shared genetic architecture and behavior traits are also genetic related to their families. However, the genetic relatedness between physiological traits and behavioral traits is not significant. For example, the muscle trait \emph {TA} is closely related to other muscle traits but shares less common genetic variants with behavior traits, which is similarly reported in \cite{wang2021unified}. Similarly, we note that the prepulse inhibition traits have a significant genetic correlation with other behavior traits, but are not significantly related to physiological traits. Specially, we notice that the prepulse inhibition trait \emph{PPlavg} that is the average of the
	inhibition intensity taken as the ratio of the prepulse response across all amplitudes shows a strong negative relationship with other traits except for several congeneric traits. In addition, \emph{pp3PPlavg} and \emph{pp6PPlavg} which are the average of the inhibition intensity during the 3-dB and 6-dB prepulse trials to the pulse-alone startle amplitude show similar characteristics.
	
%	\begin{figure}[!h]
%		\centering
%		\includegraphics[scale=0.6]{heatmap_1206.jpg}
%		\caption{Heatmap of phenotypic correlation for 24 phenotypes}
%	\end{figure}

\section{Conclusions and discussions}
\label{section:sec7}
	
In this paper, we propose semiparametric efficient estimators of the genetic covariance and genetic correlation for both continuous and discrete responses. We first derive the efficient influence functions of genetic relatedness. We propose a consistent estimator of the genetic covariance as long as one of genetic values is consistently estimated. Based on the efficient influence function, our proposed estimators are semiparametric efficient without the risk of model misspecification. Our procedures allow the data of two traits be collected from two different groups of individuals.
	
	There are also some future possible topics. For example, how to model the nonlinear correlated effect. In this paper, the genetic covariance can be zero while $EY|X$ and $EZ|X$ are nonlinearly dependent. Moreover, when $Y$ and $Z$ are multivariate, how to define a suitable measure and make inference is also of great interest. We will investigate these issues in near future.

\section*{Appendix}
\appendix
\label{appn}

\setcounter{equation}{0}
\renewcommand{\theequation}{A\arabic{equation}}
 %\setcounter{equation}{0}

%\section{Proof of theoretical results}

%\subsection*{Notation}
{\emph{Notation}:} For two positive sequences $a_n$ and $b_n$, $a_n\lesssim b_n$ means $a_n\leq Cb_n$ for all $n$, $a_n \gtrsim b_n$ if $b_n\lesssim a_n$ and $a_n \asymp b_n$ if $a_n\lesssim b_n$ and $a_n \gtrsim b_n$. $C$ is used to denote generic positive constants that may vary from place to place. What's more, we denote $a_n\lesssim_{P} b_n$ to represent $a_n\lesssim b_n$ in probability.

%\subsection*{Proof of Theorem \ref{theorem_eif}}
{\emph{Proof of Theorem \ref{theorem_eif}}:} Denote $\mathcal{P}$ as the distribution of $O=(X^{\top},Y^*,Z^*,T_y,T_z)^{\top}$. For better illustration, we can rewrite $I$ as
\begin{align*}
	I=\Psi(\mathcal{P})=E_{\mathcal{P}}\{[E_{\mathcal{P}}(Y|X)-E_{\mathcal{P}}(Y)][E_{\mathcal{P}}(Z|X)-E_{\mathcal{P}}(Z)]\}.
\end{align*}
Note that the indicator variables $T_y$ and $T_z$ are independent of $(X,Y,Z)$. Thus we have $E_{\mathcal{P}}(Y|X)=E_{\mathcal{P}}(Y^*|X,T_y=1)$ and $E_{\mathcal{P}}(Y)=E_{\mathcal{P}}(Y^*|T_y=1)$. The case for $Z$ is similar. Thus, it follows that
\begin{align*}
	I=\Psi(\mathcal{P})=E_{\mathcal{P}}\{[E_{\mathcal{P}}(Y^*|X,T_y=1)-E_{\mathcal{P}}(Y^*|T_y=1)][E_{\mathcal{P}}(Z^*|X, T_z=1)-E_{\mathcal{P}}(Z^*|T_z=1)]\}.
\end{align*}

Consider the following parametric submodel indexed by $t$, i.e.
\begin{align*}
	\mathcal{P}_t=t\tilde{ \mathcal{P}}+(1-t)\mathcal{P},
\end{align*}
where $t\in[0,1]$, and $\tilde{ \mathcal{P}}$ is a point mass at a single observation $\tilde o=(\tilde x^{\top},\tilde y^*,\tilde z^*,\tilde t  _y,\tilde t_z)^{\top}$. As mentioned in \cite{hines2022}, the efficient influence function (EIF) for $I$ at observation $\tilde o$ directly as
\begin{align*}
	\phi(\tilde o,\mathcal{P})=\frac{d\Psi(\mathcal{P}_t)}{dt}\Bigg|_{t=0},
\end{align*}
where $\Psi(\mathcal{P}_t)=E_{\mathcal{P}_t}\{[E_{\mathcal{P}_t}(Y^*|X,T_y=1)-E_{\mathcal{P}_t}(Y^*|T_y=1)][E_{\mathcal{P}_t}(Z^*|X,T_z=1)-E_{\mathcal{P}_t}(Z^*|T_z=1)]\}.$

Denote $ m^*_t(X)=E_{\mathcal{P}_t}\{[E_{\mathcal{P}_t}(Y^*|X,T_y=1)-E_{\mathcal{P}_t}(Y^*|T_y=1)]$ and $h^*_t(X)=[E_{\mathcal{P}_t}(Z^*|X,T_z=1)-E_{\mathcal{P}_t}(Z^*|T_z=1)]$. Further, the operator, $\partial_t$, applied to an arbitrary function $g(t)$, is defined as
$$\partial_t g(t)=\frac{d g(t)}{d t}\big |_{t=0}.$$
Simple calculation entails that
$$\partial_t \Psi(\mathcal{P}_t)=E_{\mathcal{P}}\{  \partial_t[m^*_t(X)  h^*_t(X)]\}+m_0^*(\tilde x)h_0^*(\tilde x)-\Psi(\mathcal{P}).$$
By the facts that
$$\partial_t E_{\mathcal{P}_t}(Y^*|X,T_y=1)=\frac{\mathbb{I}_{(\tilde x,\tilde t_y)}(X,T_y=1)}{f(X,T_y=1)}[\tilde y^*-E(Y^*|X,T_y=1)],$$
and  $$\partial_t E_{\mathcal{P}_t}(Y|T_y=1)=\frac{\mathbb{I}_{\tilde t _y}(T_y=1)}{\Pr(T_y=1)}[\tilde y^*-E(Y^*|T_y=1)],$$
we have
\begin{align*}
	E_{\mathcal{P}}\{  \partial_t[m^*_t(X)] h^*_0(X)\} &=\int \frac{\mathbb{I}_{(\tilde x,\tilde t _y)}(X,T_y=1)}{f(X,T_y=1)}[\tilde y^*-E(Y^*|X,T_y=1)]h_0^*(X) f(X)dX \\
	&=\frac{\mathbb{I}_{\tilde t _y}(T_y=1)}{\Pr(T_y=1|\tilde x)} [\tilde y^*-E(Y^*|\tilde x,T_y=1)]h_0^*(\tilde x),
\end{align*}
where the first equation holds because $E[h_0^*(X)]=0$.
Note that $T_y\perp X$, we have $\Pr(T_y=1|X)=\Pr(T_y=1),$ which implies that
\begin{align*}
	E_{\mathcal{P}}\{  \partial_t[m^*_t(X)] h^*_0(X)\} =\frac{\mathbb{I}_{\tilde t_y}(T_y=1)}{\Pr(T_y=1)} [\tilde y^*-E(Y^*|\tilde x,T_y=1)]h_0^*(\tilde x).
\end{align*}
Hence, by the chain rule and the quotient rule for derivatives, we obtain
\begin{align*}
	E_{\mathcal{P}}\{  \partial_t[m^*_t(X)  h^*_t(X)]\}=& \ \frac{\mathbb{I}_{\tilde t _y}(T_y=1)}{\Pr(T_y=1)} [\tilde y^*-E(Y^*|\tilde x,T_y=1)][E(Z^*|\tilde x,T_z=1)-E(Z^*|T_z=1)]\\
	&+\frac{\mathbb{I}_{\tilde t _z}(T_z=1)}{\Pr(T_z=1)} [\tilde z-E(Z^*|\tilde x,T_z=1)][E(Y^*|\tilde x,T_y=1)-E(Y^*|T_y=1)].
\end{align*}
{Note that $m_0^*(\tilde x)h_0^*(\tilde x) = [E(Y^*|\tilde x,T_y=1)-E(Y^*|T_y=1)][E(Z^*|\tilde x,T_z=1)-E(Z^*|T_z=1)]$.}
It implies that
\begin{align}
	\label{eif_pi}
	\begin{aligned}
		\phi(\tilde o,\mathcal{P})=&\ \frac{\mathbb{I}_{\tilde t _y}(T_y=1)}{\Pr(T_y=1)} [\tilde y^*-E(Y^*|\tilde x,T_y=1)][E(Z^*|\tilde x,T_z=1)-E(Z^*|T_z=1)]\\
		&+\frac{\mathbb{I}_{\tilde t _z}(T_z=1)}{\Pr(T_z=1)} [\tilde z^*-E(Z^*|\tilde x,T_z=1)][E(Y^*|\tilde x,T_y=1)-E(Y^*|T_y=1)]\\
		&+[E(Y^*|\tilde x,T_y=1)-E(Y^*|T_y=1)][E(Z^*|\tilde x,T_z=1)-E(Z^*|T_z=1)] -\Psi(\mathcal{P}).		
	\end{aligned}
\end{align}
Since $\phi(O,\mathcal{P})$ has finite variance, we conclude that the EIF for $I$ is
% $$EIF=\phi(\tilde o,\mathcal{P})=[\tilde y-E(Y|\tilde x)][E(Z|X)-E(Z)]+[E(Y|X)-E(Y)][\tilde z-E(Z|\tilde x)]+m_0^*(\tilde x)h_0^*(\tilde x)-\Psi(\mathcal{P}).$$
\begin{align*}
	\phi(O,\mathcal{P})=&\ \frac{\mathbb{I}_{T_y}(1)}{\Pr(T_y=1)} [Y^*-E(Y^*|X,T_y=1)][E(Z^*|X,T_z=1)-E(Z^*|T_z=1)]\\
	&+\frac{\mathbb{I}_{T_z}(1)}{\Pr(T_z=1)} [Z^*-E(Z^*|X,T_z=1)][E(Y^*|X,T_y=1)-E(Y^*|T_y=1)]\\
	&+[E(Y^*|X,T_y=1)-E(Y^*|T_y=1)][E(Z^*|X,T_z=1)-E(Z^*|T_z=1)] -\Psi(\mathcal{P}).
\end{align*}

\vspace{3mm}

{\emph{Proof of Theorem \ref{theorem_con}}:} Let $\epsilon=Y-m(X)$ and $\eta=Z-h(X)$. Note that
\begin{align*}
	G_1&=\frac{1}{N_y}\sum_{i=1}^{N_y}[Y_i-\hat m(X_i)][\hat h(X_i)-\bar Z_N]\\
	&=\frac{1}{N_y}\sum_{i=1}^{N_y}[\epsilon_i+m(X_i)-\tilde m(X_i)+\tilde m(X_i)-\hat m(X_i)][\hat h(X_i)-\tilde h(X_i)+\tilde h(X_i)-EZ+EZ-\bar Z_N]\\
	&=\frac{1}{N_y}\sum_{i=1}^{N_y}[\epsilon_i+m(X_i)-\tilde m(X_i)][\tilde h(X_i)-EZ]+\frac{1}{N_y}\sum_{i=1}^{N_y}[\tilde m(X_i)-\hat m(X_i)][\tilde h(X_i)-EZ]\\
	&\quad +\frac{1}{N_y}\sum_{i=1}^{N_y}\epsilon_i[\hat h(X_i)-\tilde h(X_i)]+\frac{1}{N_y}\sum_{i=1}^{N_y}[m(X_i)-\tilde m(X_i)][\hat h(X_i)-\tilde h(X_i)]\\
	&\quad +\frac{1}{N_y}\sum_{i=1}^{N_y}[\tilde m(X_i)-\hat m(X_i)][\hat h(X_i)-\tilde h(X_i)]+[EZ-\bar Z_N]\frac{1}{N_y}\sum_{i=1}^{N_y}[\epsilon_i+m(X_i)-\tilde m(X_i)] \\
	&\quad +[EZ-\bar Z_N]\frac{1}{N_y}\sum_{i=1}^{N_y}[\tilde m(X_i)-\hat m(X_i)]:=\sum_{i=1}^8G_{1i}.
\end{align*}
For the term $G_{12}$, by Cauchy-Schwartz inequality, we have
\begin{align*}
	&\frac{1}{N_y}\sum_{i=1}^{N_y}[\tilde m(X_i)-\hat m(X_i)][\tilde h(X_i)-EZ]\\
	\leq & \left(\frac{1}{N_y}\sum_{i=1}^{N_y}[\tilde m(X_i)-\hat m(X_i)]^2\right)^{1/2}   \left(\frac{1}{N_y}\sum_{i=1}^{N_y} [\tilde h(X_i)-EZ]^2\right)^{1/2}=o_p(1),
\end{align*}
where the last equation holds under the conditions $E[\tilde h(X_i)-EZ]^2<\infty$ and $E[\hat h(X_i)-h(X_i)]^2=o(1)$.
Similarly, due to the consistencies of $\hat m(X_i),\hat h(X_i)$ to $\tilde m(X_i), \tilde h(X_i)$, respectively, we conclude that
\begin{align*}
	G_1&=\frac{1}{N_y}\sum_{i=1}^{N_y}[\epsilon_i+m(X_i)-\tilde m(X_i)][\tilde h(X_i)-EZ]+o_p(1)\\
	&=E\{[m(X)-\tilde m(X)][\tilde h(X)-EZ]\}+o_p(1),
\end{align*}
The last equation follows from the fact that $E\{\epsilon[\tilde h(X_i)-EZ]\}=0.$ Similarly for the second term of $\tilde I_N$, we have ${G_2 = E\{[\tilde m(X) - EY][ h(X)-\tilde h(X)]\}+o_p(1).}$

For the third term of $\tilde{I}_{N}$, note that
\begin{align*}
	G_3 &= \frac{1}{N}\sum_{i=1}^{N}[\hat m(X_i) -\bar Y_N][\hat h(X_i) - \bar Z_N] \\
	&= \frac{1}{N}\sum_{i=1}^{N}[\hat m(X_i) - \tilde m(X_i) + \tilde m(X_i) - EY + EY - \bar Y_N][\hat h(X_i) - \tilde h(X_i) + \tilde h(X_i) - EZ + EZ - \bar Z_N]\\
	&= \frac{1}{N}\sum_{i=1}^{N}[\tilde m(X_i) - EY][\tilde h(X_i) - EZ] + o_p(1)\\
	&= E{[\tilde m(X) - EY][\tilde h(X) - EZ]} + o_p(1).
\end{align*}
The third equation holds due to the consistencies of the estimators.

% \begin{align*}
% G_2&=\frac{1}{N_z}\sum_{i=1}^{N_z}[\hat m(X_i)-\tilde m(X_i)+\tilde m(X_i)-EY+EY-\bar Y_N][\eta_i+h(X_i)-EZ+EZ-\bar Z_N]\\
% &=\frac{1}{N_z}\sum_{i=1}^{N_z}[\tilde m(X_i)-EY][h(X_i)-EZ]+o_p(1)\\
% &=E\{[\tilde m(X)-EY][h(X)-EZ]\}+o_p(1).
% \end{align*}
In sum, we have
\begin{align*}
	\tilde I_N ={G_1 + G_2 + G_3} = E[\tilde m(X_i)-m(X_i)][h(X_i)-\tilde h(X_i)]+I+o_p(1).
\end{align*}
Then the Theorem follows.

\vspace{3mm}

{\emph{Proof of Theorem \ref{theorem_asy1}};}
Note that the proof of this theorem is a special case of that of Theorem \ref{theorem_asy2} with $g_1(x)=g_2(x)=x$. Hence we omit the detail.

\vspace{3mm}

{\emph{Proof of Theorem \ref{theorem_ci}}:}
If $(\hat \sigma^2-\sigma^2)/\sigma^2=o_p(1)$ holds, note that $$  \frac{I_n^*-I }{\sigma} \overset{d}{\rightarrow} N\Big(0,1\Big),$$ which implies
$$  \frac{I_n^*-I }{\hat \sigma} \overset{d}{\rightarrow} N\Big(0,1\Big).$$
Hence we can finish the proof.

In the following, we aim to show that $(\hat \sigma^2-\sigma^2)/\sigma^2=o_p(1)$.
Denote
\begin{align*}
	\delta_i-n^{-1}I=&\frac{1}{n_y}\epsilon_i^*I(i\in\D_{2y})+\frac{1}{n_z}\eta_i^*I(i\in\D_{2z})+\frac{1}{n}\xi^*_i,
\end{align*}
and $\tilde K_1=\sum_{i\in\D_2}(\delta_i-n^{-1}I)^2$, where $ \epsilon^*_i=\epsilon_i[h(X_i)-EZ]$, $\eta^*_i=\eta_i[m(X_i)-EY]$, and $\xi_i^*=[m(X_i)-EY][ h(X_i)-EZ]-I$. Similarly, we can definite $\tilde K_2$. Denote
\begin{align*}
	\tilde {\sigma}^2=\frac{\tilde K_1+\tilde K_2}{4}.
\end{align*}

We have
\begin{align}
	\frac{\hat \sigma^2-\sigma^2}{\sigma^2}=\frac{\tilde {\sigma}^2-\sigma^2}{\sigma^2}+\frac{\hat \sigma^2-\tilde {\sigma}^2}{\sigma^2}.
\end{align}
We first consider the first term. Note that $E({\tilde{\sigma}^2-\sigma^2})=0$, and
\begin{align*}
	\mbox{Var}({\tilde{\sigma}^2-\sigma^2} )\lesssim
	\mbox{Var}(\tilde {K}_1)+\mbox{Var}(\tilde {K}_2).
\end{align*}
We have
\begin{align*}
	\mbox{Var}(\tilde {K}_1) \lesssim E[\sum_{i\in \D_2}(\delta_i-n^{-1}I)^4] &\lesssim \frac{1}{n_y^3}E(\epsilon_i^{*4})+\frac{1}{n_z^3}E(\eta_i^{*4})+\frac{1}{n^3}E(\xi_i^{*4}),
\end{align*}
and $\mbox{Var}(\tilde {K}_2)\lesssim \dfrac{1}{n_y^3}E(\epsilon_i^{*4})+\dfrac{1}{n_z^3}E(\eta_i^{*4})+\dfrac{1}{n^3}E(\xi_i^{*4}),$
which implies that
\begin{align*}
	{\tilde{\sigma}^2-\sigma^2}= O_p\Big(\frac{1}{n_y^{3/2}}+\frac{1}{n_z^{3/2}}\Big).
\end{align*}
Since
\begin{align*}
	\sigma^2=\frac{1}{2n_y}E(\epsilon_i^{*2})+\frac{1}{2n_z}{E(\eta_i^{*2})}+\frac{1}{2n}{E(\xi_i^{*2})} +\frac{N_0}{N_yN_z}E[\epsilon^*\eta^*]=O_p\left(\frac{1}{n_y}+\frac{1}{n_z}\right),
\end{align*}
It follows that
\begin{align*}
	\frac{\tilde  {\sigma}^2-\sigma^2}{\sigma^2}=o_p(1).
\end{align*}

Next, we turn to consider the bound for $(\hat \sigma^2-\tilde{\sigma}^2)/\sigma^2$. Note that
\begin{align*}
	\frac{\hat \sigma^2-\tilde{\sigma}^2}{\sigma^2}=\frac{K_1-\tilde K_1}{4\sigma^2}+\frac{K_2-\tilde K_2}{4\sigma^2}.
\end{align*}
We can rewrite $K_1-\tilde K_1$ as
\begin{align*}
	K_1-\tilde K_1&=\sum_{i\in\D_2}(\hat \delta_i-n^{-1}I_n)^2-\sum_{i\in\D_2}(\delta_i-n^{-1}I)^2\\
	&=\sum_{i\in\D_2}(\hat \delta_i-n^{-1}I_n-\delta_i+n^{-1}I)^2+2\sum_{i\in\D_2}(\delta_i-n^{-1}I)\sum_{i\in\D_2}(\hat \delta_i-n^{-1}I_n-\delta_i+n^{-1}I)\\
	&\leq \sum_{i\in\D_2}(\hat \delta_i-n^{-1}I_n-\delta_i+n^{-1}I)^2+2\sqrt{\tilde K_1}\sqrt{ \sum_{i\in\D_2}(\hat \delta_i-n^{-1}I_n-\delta_i+n^{-1}I)^2}\\
	&=:K_{11}+2K_{12}.
\end{align*}
For the term $K_{11}$, we have
\begin{align*}
	\sum_{i\in\D_2}(\hat \delta_i-n^{-1}I_n-\delta_i+n^{-1}I)^2\leq 2n^{-1}(I_n -I)^2+2 \sum_{i\in\D_2}(\hat \delta_i-\delta_i)^2.
\end{align*}
By the asymptotic normality of $I_n$, we have $(I_n-I)^2=O_p(2\sigma^2)$, which implies that $n^{-1}(I_n -I)^2/4\sigma^2=o_p(1)$.

Hence, it is left to show that $\sum_{i\in\D_2}(\hat \delta_i-\delta_i)^2/4\sigma^2=o_p(1).$ Note that
\begin{align*}
	\sum_{i\in \D_2}(\hat\delta_i-\delta_i)^2\lesssim &\ \frac{1}{n^2_y}\sum_{i\in \D_{2y}}\{[Y_i-\hat m_{\D_{1}}(X_i)][\hat h_{\D_1}(X_i)-\bar Z_n]-[Y_i- m(X_i)][h(X_i)-EZ]\}^2\\
	&+\frac{1}{n_z^2}\sum_{i\in \D_{2z}}\{[\hat m_{\D_1}(X_i)-\bar Y_n][Z_i- \hat h_{\D_1}(X_i)]-[m(X_i)-EY][Z_i-   h(X_i)]\}^2\\
	&+\frac{1}{n^2}\sum_{i\in \D_2}\{[\hat m_{\D_1}(X_i)-\bar Y_n][ \hat h_{\D_1}(X_i)-\bar Z_n]-[m(X_i)-EY][ h(X_i)-EZ]\}^2\\
	=:& \ K_{111}+K_{112}+K_{113}.
\end{align*}
We first consider the bound of the term $K_{111}$. Note that
\begin{align*}
	K_{111}&=\frac{1}{n^2_y}\sum_{i\in \D_{2y}}\{[Y_i-\hat m_{\D_{1}}(X_i)][\hat h_{\D_1}(X_i)-\bar Z_n]-[Y_i- m(X_i)][h(X_i)-EZ]\}^2 \\
	&\lesssim_P \frac{1}{n^2_y}\sum_{i\in \D_{2y}}\epsilon_i^2[\hat h_{\D_1}(X_i)-h(X_i)]^2 +[EZ-\bar Z_n]^2 \frac{1}{n^2_y}\sum_{i\in \D_{2y}}\epsilon^2_i\\
	&\quad +\frac{1}{n^2_y}\sum_{i\in \D_{2y}}[\hat m_{\D_{1}}(X_i)-m(X_i)]^2[h(X_i)-EZ]^2 +\frac{1}{n^2_y}\sum_{i\in \D_{2y}}[\hat m_{\D_{1}}(X_i)-m(X_i)]^2[\hat h_{\D_1}(X_i)-h(X_i)]^2 \\
	&\quad +[EZ-\bar Z_n]^2\frac{1}{n^2_y}\sum_{i\in \D_{2y}}[\hat m_{\D_{1}}(X_i)-m(X_i)]^2.
\end{align*}

Based on the fact that $[EZ-\bar Z_n]^2=O_p(n_z^{-1})$ and under Conditions \ref{condition1}--\ref{condition4}, it is easily to show that
\begin{align*}
	E(|K_{111}|)=o_p\left(\frac{1}{n_y}\right),
\end{align*}
which follows that $K_{111}=o_p\left({n_y}^{-1}\right)$.

Similarly, we can obtain that
\begin{align*}
	K_{112}=o_p\left(\frac{1}{n_z}\right), \text{ and } K_{113}=o_p\left(\frac{1}{n}\right).
\end{align*}
It follows that
\begin{align*}
	\frac{\sum_{i\in\D_2}(\hat \delta_i-\delta_i)^2}{4\sigma^2}=o_p(1),
\end{align*}
which implies that $K_{11}/4\sigma^2=o_p(1).$

Similarly as the discussion of $\tilde \sigma^2$, we can show that
\begin{align*}
	\frac{\tilde K_1}{4\sigma^2}=o_p(1).
\end{align*}
Hence, for the term $K_{12}$, we have
\begin{align*}
	\frac{K_{12}}{4\sigma^2}&=\sqrt{\frac{\tilde K_1}{4\sigma^2}} \sqrt{\frac{K_{11}}{4\sigma^2}}=o_p(1).
\end{align*}

Based on the above results, it follows that $$(K_1-\tilde K_1)/4\sigma^2=o_p(1).$$
Similarly, we can show that
$$(K_2-\tilde K_2)/4\sigma^2=o_p(1).$$
So that we finish the proof.

\vspace{3mm}

{\emph{Proof of Proposition \ref{theorem_eifp}}:}
Similarly, we can rewrite $\rho$ as $$\rho=\Psi_1(\mathcal{P})=\frac{Cov[m(X),h(X)]}{\sqrt{Var[m(X)]Var[h(X)]}}.$$ Consider the same parametric submodel,
\begin{align*}
	\mathcal{P}_t=t\tilde{ \mathcal{P}}+(1-t)\mathcal{P},
\end{align*}
where $t\in[0,1]$. Hence, the efficient influence function(EIF) for $\rho$ at observation $\tilde o$ directly as
\begin{align*}
	\phi_1(\tilde o,\mathcal{P})=\frac{d\Psi_1(\mathcal{P}_t)}{dt}|_{t=0}=\frac{\phi(\tilde o,\mathcal{P})}{\sqrt{B_0^yB_0^z}}-\frac{\Psi_1(\mathcal{P})}{2B_0^yB_0^z}
	\partial_t(B_t^yB_t^u),
\end{align*}
where $\phi(\tilde o,\mathcal{P})$ is the EIF for $I$,  $B_t^y=E_{\mathcal{P}_t}\{[E_{\mathcal{P}_t}(Y^*|X,T_y=1)-E_{\mathcal{P}_t}(Y^*|T_y=1)]^2\}$, and $B_0^y=Var(E[Y^*|X,T_y=1])$, and the definition of $B_t^z$ is similar. Similarly, as the arguments in Theorem \ref{theorem_con}, we can show that $\partial_t B_t^u$ equals to
\begin{align*}
	\phi_u(\tilde o,\mathcal{P})=&\ 2\frac{\mathbb{I}_{\tilde t_u}(T_u=1)}{P(T_u=1)} [\tilde u^*-E(U^*|\tilde x,T_u=1)][E(U^*|\tilde x,T_u=1)-E(U^*|T_u=1)]\\
	&+[E(U^*|\tilde x,T_u=1)-E(U^*|T_u=1)]^2-B_0^u.
\end{align*}
By the chain rule and the quotient rule for derivatives, we obtain
\begin{align*}
	\phi_1(\tilde o,\mathcal{P})
	&=\frac{\phi(\tilde o,\mathcal{P})}{\sqrt{B_0^yB_0^z}}-\frac{\Psi_1(\mathcal{P})}{2B_0^yB_0^z}[\phi_y(\tilde o,\mathcal{P})B_0^z+B_0^y\phi_z(\tilde o,\mathcal{P})]\\
	&=\frac{S_{yz}}{\sqrt{B_0^yB_0^z}}-\rho\frac{S_{yy}}{2B_0^y}-\rho\frac{S_{zz}}{2B_0^z}.
\end{align*}
Note that the variance of $\phi_1(O,\mathcal{P})$ is finite, which implies that the EIF for $\rho$ is  $\phi_1(O,\mathcal{P})$.

\vspace{3mm}

{\emph{Proof of Theorem \ref{theorem_asyp}}:}
Before giving the proof, a necessary lemma is presented.

\begin{lemma}\label{lemma_var}
	Suppose Conditions \ref{condition1}--\ref{condition2} are satisfied. The asymptotically linear representation for $\hat B_{0}^y$ and $\hat B_{0}^z$ respectively are
	\begin{align*}
		&\hat B_0^y-B_0^y=\sum_{i\in\D_{2}}S^2_{yy,i}+o_p\left(\frac{1}{\sqrt{n}}+\frac{1}{\sqrt{n_y}}\right),\\
		&\hat B_0^z-B_0^z =\sum_{i\in\D_{2}}S^2_{zz,i}+o_p\left(\frac{1}{\sqrt{n}}+\frac{1}{\sqrt{n_z}}\right).
	\end{align*}
	Here  $$S^2_{yy,i}=\frac{2}{n_y}\epsilon_i[h(X_i)-EZ]\mathbb{I}(T_{y,i}=1)+\frac{1}{n}\{[m(X_i)-EY ]^2-B_0^y\},$$
	and $$S^2_{zz,i}=\frac{2}{n_z}\eta_i[m(X_i)-EY]\mathbb{I}(T_{z,i}=1)+\frac{1}{n}\{[h(X_i)-EZ]^2-B_0^z\}.$$
\end{lemma}
\begin{proof}
	The proof of this lemma is similar to that of Theorem \ref{theorem_asy1} and thus omitted here.
\end{proof}

Note that
\begin{align*}
	\rho_n-\rho&=\Big (\frac{I_n}{\sqrt{\hat B_0^y \hat B_0^z}}-\frac{I}{\sqrt{B_0^y   B_0^z}}\Big)\\
	&=\frac{ (I_n-I)}{\sqrt{B_0^y  B_0^z}}+ I \Big(\frac{1}{\sqrt{\hat B_0^y \hat B_0^z} }-\frac{1}{\sqrt{  B_0^y   B_0^z} }\Big)+ (I_n-I) \Big(\frac{1}{\sqrt{\hat B_0^y \hat B_0^z} }-\frac{1}{\sqrt{  B_0^y   B_0^z} }\Big)\\
	&=:C_1+C_2+C_3.
\end{align*}
According to the proof of Theorem~\ref{theorem_asy1}, we have
\begin{align*}
	C_1=\sum_{i\in\D_2}\frac{S_{yz,i}}{\sqrt{B_0^yB_0^z}}+o_p\left(\frac{1}{\sqrt{n_z}}+\frac{1}{\sqrt{n_y}}\right),
\end{align*}
where
\begin{align*}
	S_{yz,i}=\frac{1}{n_y}\epsilon_i[h(X_i)-EZ]\mathbb{I}(T_{y,i}=1)+\frac{1}{n_z}\eta_i[m(X_i)-EY]I(D_{z,i}=1)+\frac{1}{n}[m(X_i)-EY ][h(X_i)-EZ].
\end{align*}
Through the Taylor series expansion, we have
\begin{align*}
	\frac{1}{\sqrt{\hat B_0^y \hat B_0^z} }-\frac{1}{\sqrt{  B_0^y   B_0^z} }=&\ -\frac{1}{2}(  B_0^y   B_0^z)^{-3/2}B_0^z(\hat B_0^y-B_0^y)-\frac{1}{2}(  B_0^y   B_0^z)^{-3/2}B_0^y(\hat B_0^z-B_0^z)\\
	&+o_p(|\hat B_0^y-B_0^y|^2+|\hat B_0^z-B_0^z|^2).
\end{align*}

From Lemma~\ref{lemma_var}, it follows that
\begin{align*}
	\frac{1}{\sqrt{\hat B_0^y \hat B_0^z} }-\frac{1}{\sqrt{  B_0^y   B_0^z} }& =-\frac{1}{2}(  B_0^y   B_0^z)^{-3/2}B_0^z (\hat B_0^y-B_0^y)-\frac{1}{2}(  B_0^y   B_0^z)^{-3/2}B_0^y (\hat B_0^z-B_0^z)+o_p\left( \frac{1}{\sqrt{n_z}}+\frac{1}{\sqrt{n_y}}\right)\\
	&=-\frac{(  B_0^y   B_0^z)^{-3/2}}{2}\sum_{i\in \mathcal{D}_2}\{B_0^zS_{yy,i}+B_0^yS_{zz,i}\}+o_p\left( \frac{1}{\sqrt{n_z}}+\frac{1}{\sqrt{n_y}}\right).
\end{align*}
Thus,
\begin{align*}
	C_2& =-\frac{\rho}{2{B_0^y   B_0^z}} \sum_{i\in \mathcal{D}_2}\{B_0^zS_{yy,i}+B_0^yS_{zz,i}\}+o_p\left( \frac{1}{\sqrt{n_z}}+\frac{1}{\sqrt{n_y}}\right).
\end{align*}

Also note that $I_n-I=o_p\left(1\right)$. Hence it can be easily obtained that $C_3=o_p\left( \dfrac{1}{\sqrt{n_z}}+\dfrac{1}{\sqrt{n_y}}\right)$.

Combining the above results, we then conclude that
\begin{align*}
	\rho_n-\rho= \sum_{i\in\mathcal{D}_2} \frac{S_{yz,i}}{\sqrt{B_0^yB_0^z}}-\rho\frac{S_{yy,i}}{2B_0^y}-\rho\frac{S_{zz,i}}{2B_0^z}  +o_p\left( \frac{1}{\sqrt{n_z}}+\frac{1}{\sqrt{n_y}}\right).
\end{align*}
Similarly for $\rho_n'$, we have
\begin{align*}
	\rho_n'-\rho = \sum_{i\in\mathcal{D}_1} \frac{S_{yz,i}}{\sqrt{B_0^yB_0^z}}-\rho\frac{S_{yy,i}}{2B_0^y}-\rho\frac{S_{zz,i}}{2B_0^z} +o_p\left( \frac{1}{\sqrt{n_z}}+\frac{1}{\sqrt{n_y}}\right).
\end{align*}

Clearly, $\rho_n$ and $\rho_n'$ are asymptotically independent. Then we conclude that
\begin{align*}
	\rho_n^*-\rho = \sum_{i\in\mathcal{D}} \frac{S^*_{yz,i}}{\sqrt{B_0^yB_0^z}}-\rho\frac{S^*_{yy,i}}{2B_0^y}-\rho\frac{S^*_{zz,i}}{2B_0^z} +o_p\left( \frac{1}{\sqrt{N_z}}+\frac{1}{\sqrt{N_y}}\right).
\end{align*}
Here
\begin{align*}
	S^*_{yz,i}=&\ \frac{1}{N_y}\epsilon_i[h(X_i)-EZ]\mathbb{I}(T_{y,i}=1)+\frac{1}{N_z}\eta_i[m(X_i)-EY]I(D_{z,i}=1)\\
	&+\frac{1}{N}\{[m(X_i)-EY ][h(X_i)-EZ]-I\},
\end{align*}
and $S^*_{yy,i}$ and $S^*_{zz,i}$ are similarly defined.

\vspace{3mm}

{\emph{Proof of Proposition~\ref{theorem_eif2}}:}
We first rewrite $I$ as $$I=\Psi(\mathcal{P})=
E_{\mathcal{P}}\{g^*_1(X) g^*_2(X) \},$$
where  $g^*_{1}(X)=g_1[E_{\mathcal{P}}(Y^*|X,T_y=1)]-E_{\mathcal{P}}\{g_1[E_{\mathcal{P}}(Y^*|X,T_y=1)]\}$ and $g^*_{2}(X)=g_2[E_{\mathcal{P}}(Z^*|X,T_z=1)]-E_{\mathcal{P}}\{g_2[E_{\mathcal{P}}(Z^*|X,T_z=1)]\}$.
Consider the same parametric submodel indexed by $t$, i.e. $$\mathcal{P}_t=t\tilde{ \mathcal{P}}+(1-t)\mathcal{P},$$
where $t\in[0,1]$, and $\tilde{ \mathcal{P}}$ is a point mass at a single observation $\tilde o$.
So that the efficient influence function (EIF) for $I$ at observation $\tilde o$ directly as
\begin{align*}
	\phi(\tilde o,\mathcal{P})=\frac{d\Psi(\mathcal{P}_t)}{dt}|_{t=0}= E_{\mathcal{P}}\{\partial_t[g^*_{1t}(X)g^*_{2t}(X)]\}+\partial_tE_{\mathcal{P}_t}\{g^*_{1}(X)g^*_{2}(X)\}.
\end{align*}
where $g^*_{1t}(X)=g_1[E_{\mathcal{P}_t}(Y^*|X,T_y=1)]-E_{\mathcal{P}_t}\{g_1[E_{\mathcal{P}_t}(Y^*|X,T_y=1)]\}$ and $g^*_{2t}(X)=g_2[E_{\mathcal{P}_t}(Z^*|X,T_z=1)]-E_{\mathcal{P}_t}\{g_2[E_{\mathcal{P}_t}(Z^*|X,T_z=1)]\}$.

Firstly, we focus on the term $\partial_t g_{1t}^*$. Note that
\begin{align*}
	\partial_t g_{1t}^*(X)&=\partial_t \Big[g_1[E_{\mathcal{P}_t}(Y^*|X,T_y=1)]\Big]-\partial_t\Big(\int  g_1[E_{\mathcal{P}_t}(Y^*|X,T_y=1)] f_t(X)dX\Big)\\
	&=g'_1[E_{\mathcal{P}}(Y^*|X,T_y=1)]\partial_t E_{\mathcal{P}_t}(Y^*|X,T_y=1)]-  g_1[E_{\mathcal{P}}(Y^*|\tilde x,T_y=1)]\\
	&\quad +E_{\mathcal{P} }(g_1[E_{\mathcal{P} }(Y^*|X,T_y=1)])-E_{\mathcal{P}}\left(\partial_t \Big[g_1[E_{\mathcal{P}_t}(Y^*|X,T_y=1)]\Big]\right)=:\psi_1(\tilde o,\mathcal{P}).
\end{align*}
Similarly, we can get
\begin{align*}
	\partial_t g_{2t}^*(X)&=\partial_t \Big[g_2[E_{\mathcal{P}_t}(Z^*|X,T_z=1)]\Big]-\partial_t\Big(\int  g_2[E_{\mathcal{P}_t}(Z^*|X,T_z=1)] f_t(X)dX\Big)\\
	&=g'_2[E_{\mathcal{P}}(Z^*|X,T_z=1)]\partial_t E_{\mathcal{P}_t}(Z^*|X,T_z=1)]-  g_2[E_{\mathcal{P}}(Z^*|\tilde x,T_z=1)]\\
	&\quad +E_{\mathcal{P} }(g_2[E_{\mathcal{P} }(Z^*|X,T_z=1)])-E_{\mathcal{P}}\left(\partial_t \Big[g_1[E_{\mathcal{P}_t}(Z^*|X,T_z=1)]\Big]\right)=:\psi_2(\tilde o,\mathcal{P}).
\end{align*}
By the chain rule and the quotient rule for derivatives, it follows that
\begin{align*}
	E_{\mathcal{P}}\{\partial_t[g^*_{1t}(X)g^*_{2t}(X)]\}&= E_{\mathcal{P}}\{\psi_1(\tilde o,\mathcal{P})g^*_{2}(X)\}+ E_{\mathcal{P}}\{ g^*_{1}(X)\psi_2(\tilde o,\mathcal{P})\}\\
	&=E_{\mathcal{P}}\{ g'_1[E_{\mathcal{P}}(Y^*|X,T_y=1)]\partial_t E_{\mathcal{P}_t}(Y^*|X,T_y=1)] g^*_{2}(X)\}\\
	&\quad + E_{\mathcal{P}}\{ g^*_{1}(X)g'_2[E_{\mathcal{P}}(Z^*|X,T_z=1)]\partial_t E_{\mathcal{P}_t}(Z^*|X,D_z)]\}.
\end{align*}
The last equality holds because $E_{\mathcal{P}}[g_i^*(X)]=0$, $i=1,2$.
According to Theorem~\ref{theorem_eif}, we have
\begin{align*}
	\partial_t E_{\mathcal{P}_t}(Y^*|X,T_y=1)=\frac{\mathbb{I}_{(\tilde x,\tilde t _y)}(X,T_y=1)}{f(X,T_y=1)}[\tilde y^*-E(Y^*|X,T_y=1)],
\end{align*}
which implies that
\begin{align*}
	&E_{\mathcal{P}}\{ g'_1[E_{\mathcal{P}_t}(Y^*|X,T_y=1)]\partial_t E_{\mathcal{P}_t}(Y^*|X,T_y=1)] g^*_{2}(X)\}\\
	=& \frac{\mathbb{I}_{ \tilde t _y}(T_y=1)}{P(T_y=1)}g'_1[E_{\mathcal{P}}(Y^*|\tilde x,T_y=1)][\tilde y^*-E(Y^*|\tilde x,T_y=1)] g^*_{2}(\tilde x),
\end{align*}
where the last equation holds because $T_y\perp X.$

Similarly, we have
\begin{align*}
	&E_{\mathcal{P}}\{ g'_2[E_{\mathcal{P}_t}(Z^*|X,T_z=1)]\partial_t E_{\mathcal{P}_t}(Z^*|X,T_z=1)] g^*_{1}(X)\}\\
	=& \frac{\mathbb{I}_{ \tilde t _z}(T_z=1)}{P(T_z=1)}g'_2[E_{\mathcal{P}}(Z^*|\tilde x,T_z=1)][\tilde z^*-E(Z^*|\tilde x,T_z=1)] g^*_{1}(\tilde x).
\end{align*}

Further, we have
\begin{align*}
	\partial_tE_{\mathcal{P}_t}\{g^*_{1}(X)g^*_{2}(X)\}=g^*_{1}(\tilde x)g^*_{2}(\tilde x)-\Psi(\mathcal{P}).
\end{align*}
Based on the above results, we obtain
\begin{align*}
	\phi(\tilde o,\mathcal{P})=&\ \frac{d\Psi(\mathcal{P}_t)}{dt}|_{t=0}\\
	=&\ \frac{\mathbb{I}_{ \tilde t _y}(T_y=1)}{P(T_y=1)}g'_1[E_{\mathcal{P}}(Y^*|\tilde x,T_y=1)][\tilde y^*-E(Y^*|\tilde x,T_y=1)] g^*_{2}(\tilde x) \\
	&+\frac{\mathbb{I}_{ \tilde t _z}(T_z=1)}{P(T_z=1)}g'_2[E_{\mathcal{P}}(Z^*|\tilde x,T_z=1)][\tilde z^*-E(Z^*|\tilde x,T_z=1)] g^*_{1}(\tilde x)\\
	&+g^*_{1}(\tilde x)g^*_{2}(\tilde x)-\Psi(\mathcal{P}).
\end{align*}
Note that the variance of $\phi(O,\mathcal{P})$ is finite, which implies that the EIF for $I$ is  $\phi(O,\mathcal{P})$.

Given the EIF of $I$, we can get the results of $\rho$ by using the similar arguments of the proof of Theorem~\ref{theorem_eifp} and thus omitted here.

\vspace{3mm}

{\emph{Proof of Theorem \ref{theorem_asy2}}:} Let
\begin{align*}
	I_n&=\frac{1}{{n_y}}\sum_{i\in\D_{2y}}
	\hat g'_{1i}[Y_i-\hat m_i][\hat g_{2i}-\bar{g}_2] +\frac{1}{ {n_z}}\sum_{i\in\D_{2z}}  \hat g_{2i}'[Z_i-\hat h_i]  [\hat g_{1i}-\bar{g}_1]+\frac{1}{{n}}\sum_{i\in\D_2}
	[\hat g_{2i}-\bar{g}_2][\hat g_{1i}-\bar{g}_1]\\
	&=:I_{n1}+I_{n2}+I_{n3}.
\end{align*}

For the term $I_{n1}$, we have
\begin{align*}
	I_{n1}&=\frac{1}{  {n_y}}\sum_{i\in\D_{2y}}
	\hat g'_{1i}[Y_i-\hat m_i][\hat g_{2i}-\bar{g}_2]\\
	&=\frac{1}{  {n_y}}\sum_{i\in\D_{2y}}
	\hat g'_{1i}[\epsilon_i+m_i-\hat m_i][\hat g_{2i}-g_{2i}+g_{2i}-Eg_2+Eg_2-\bar{g}_2]\\
	&=\frac{1}{  {n_y}}\sum_{i\in\D_{2y}}
	\hat g'_{1i}\epsilon_i(\hat g_{2i}-g_{2i})
	+\frac{1}{  {n_y}}\sum_{i\in\D_{2y}}
	\hat g'_{1i}\epsilon_i(g_{2i}-Eg_2)
	+\frac{1}{  {n_y}}\sum_{i\in\D_{2y}}
	\hat g'_{1i}\epsilon_i(Eg_2-\bar{g}_2)\\
	&\quad+\frac{1}{  {n_y}}\sum_{i\in\D_{2y}}
	\hat g'_{1i}(m_i-\hat m_i)(\hat g_{2i}-   \bar g_{2})  =:\sum_{i=1}^4 D_i.
\end{align*}
In the following part, we will show that both $D_1$ and $D_3$ are negligible, while both
$D_2$ and $D_4$ contain the leading terms.

For the term $D_1$, it can be rewritten as
\begin{align*}
	D_1=\frac{1}{  {n_y}}\sum_{i\in\D_{2y}}
	(\hat g'_{1i}-g'_{1i})\epsilon_i(\hat g_{2i}-g_{2i})+\frac{1}{  {n_y}}\sum_{i\in\D_{2y}}
	g'_{1i} \epsilon_i(\hat g_{2i}-g_{2i})=:D_{11}+D_{12}.
\end{align*}
For the first term, we have $E(D_{11})=0$, and
\begin{align*}
	E(D^2_{11}|\D_1)= \frac{1}{{n^2_y}}\sum_{i\in\D_{2y}}
	E\left[\sigma^2(X)(\hat g'_{1i}-g'_{1i})^2 (\hat g_{2i}-g_{2i})^2|\D_1\right].
\end{align*}

Under the conditions $|g'_1(m_i)-g'_1(\hat m_i)|\leq L|m_i-\hat m_i|$ and  $g'_2(\cdot)<C<\infty$, we have
\begin{align*}
	E(D^2_{11})\lesssim  \frac{1}{{n_y}}
	E\left[\sigma^2(X)(\hat m_{i}-m_{i})^2 (\hat h_{i}-h_{i})^2\right] =o\left(\frac{1}{n_y}\right),
\end{align*}
where the last equation holds under the condition $E\left[\sigma^2(X)(\hat m_{i}-m_{i})^2 (\hat h_{i}-h_{i})^2\right]=o(1)$.
Further, we can similarly show that $E(D_{12})=0$, and \begin{align*}
	E(D^2_{12})\lesssim \frac{1}{{n_y}}
	E\left[\sigma^2(X)(\hat h_{i}-h_{i})^2\right] =o\left(\frac{1}{n_y}\right),
\end{align*}
when the conditions $E[(\hat h_{i}-h_{i})^4]=o(n_y^{-1})$ and  $g'_2(\cdot)<C<\infty$ are satisfied.
It follows that
\begin{align*}
	D_1=D_{11}+D_{12}=o_p\left(\frac{1}{\sqrt{n_y}}\right).
\end{align*}

For the term $D_2$, it can be rewritten as
\begin{align*}
	D_2=\frac{1}{  {n_y}}\sum_{i\in\D_{2y}}
	(\hat g'_{1i}- g'_{1i})\epsilon_i(g_{2i}-Eg_2)+\frac{1}{  {n_y}}\sum_{i\in\D_{2y}}
	g'_{1i}\epsilon_i(g_{2i}-Eg_2)
\end{align*}
Under the conditions $|g'_1(m_i)-g'_1(\hat m_i)|\leq L|m_i-\hat m_i|$ and  $E[\sigma^2(X)(\hat m_i-m_i)^2(g_2-Eg_2)^2]=o(1)$, it can similarly obtained that
\begin{align}\label{gd2}
	D_2=\frac{1}{  {n_y}}\sum_{i\in\D_{2y}}
	g'_{1i}\epsilon_i(g_{2i}-Eg_2)+o_p\left(\frac{1}{\sqrt{n_y}}\right).
\end{align}

Now we turn to consider $Eg_2-\bar g_2$. Under conditions that $g'_2(\cdot)\leq C<\infty$ and $E[(h_i-\hat h_i)^4]=o_p(n_z^{-1})$, we have
\begin{align*}
	Eg_2-\bar{g_2}
	&=Eg_2-\frac{1}{n_z}\sum_{i\in\D_{2z}}g_2(h_i)+\frac{1}{n_z}\sum_{i\in\D_{2z}}g_2(h_i)-g_2(\hat h_i)\\
	&=Eg_2-\frac{1}{n_z}\sum_{i\in\D_{2z}}g_2(h_i)+\frac{1}{n_z}\sum_{i\in\D_{2z}}g'_2(\tilde h_i)(h_i-\hat h_i)=o_p(n_z^{-1/4}),
\end{align*}
where $\tilde h_i$ is between $h_i$ and $\hat h_i$.
Similarly, we can get $Eg_1-\bar g_1=o_p(n_y^{-1/4}).$

For the term $D_3$, it can be rewritten as
\begin{align*}
	D_3&=(Eg_2-\bar{g}_2)\frac{1}{  n_y}\sum_{i\in\D_{2y}}
	\hat g'_{1i} \epsilon_i.
\end{align*}
Given $\D_{1y}$, it can be easily to show that
$$\frac{1}{  n_y}\sum_{i\in\D_{2y}}
\hat g'_{1i} \epsilon_i=O_p\left(\frac{1}{\sqrt{n_y}}\right).$$
Also based on the fact $Eg_2-\bar g_2=o_p(n_z^{-1/4})$, we easily get that $D_3=o_p(n_y^{-1/2}).$

So far, we have
\begin{align*}
	I_{n1}& = \frac{1}{  n_y}\sum_{i\in\D_{2y}}
	g'_{1i}\epsilon_i(g_{2i}-Eg_2)+D_4 +o_p\left(\frac{1}{\sqrt{n_y}}\right).
\end{align*}
Next, we consider the bound for $D_4$. Note that
\begin{align*}
	D_4&=\frac{1}{ n_y}\sum_{i\in\D_{2y}}
	\hat g'_{1i} (m_i-\hat m_i)(\hat g_{2i}-\bar g_{2})\\
	&=  \frac{1}{  n_y}\sum_{i\in\D_{2y}}
	(\hat g'_{1i}-g'_{1i})(m_i-\hat m_i)(\hat g_{2i}-\bar g_{2})+\frac{1}{ n_y}\sum_{i\in\D_{2y}}
	g'_{1i}(m_i-\hat m_i)(\hat g_{2i}-\bar g_{2})\\
	&=:D_{41}+D_{42}.
\end{align*}
For the term $D_{41}$, we have
\begin{align*}
	D_{41}=&\  \frac{1}{ n_y}\sum_{i\in\D_{2y}}
	(\hat g'_{1i}-g'_{1i})(m_i-\hat m_i)(\hat g_{2i}-  g_{2i})+ \frac{1}{ n_y}\sum_{i\in\D_{2y}}
	(\hat g'_{1i}-g'_{1i})(m_i-\hat m_i)(  g_{2i}-E g_{2})\\
	&+(E g_{2}-\bar g_{2})  \frac{1}{ n_y}\sum_{i\in\D_{2y}}
	(\hat g'_{1i}-g'_{1i})(m_i-\hat m_i).
\end{align*}
For the first term,
\begin{align*}
	&\frac{1}{ n_y}\sum_{i\in\D_{2y}}
	(\hat g'_{1i}-g'_{1i})(m_i-\hat m_i)(\hat g_{2i}-  g_{2i})\\
	\leq &\  \left[\frac{1}{ n_y}\sum_{i\in\D_{2y}}
	(\hat g'_{1i}-g'_{1i})^2(m_i-\hat m_i)^2\right]^{1/2} \left[ \frac{1}{ n_y}\sum_{i\in\D_{2y}} (\hat g_{2i}-  g_{2i})^2\right]^{1/2}\\
	\lesssim &\  \left[\frac{1}{ n_y}\sum_{i\in\D_{2y}}
	(m_i-\hat m_i)^4\right]^{1/2} \left[ \frac{1}{ n_y}\sum_{i\in\D_{2y}} (\hat h_{i}- h_{i})^2\right]^{1/2}=o_p\left(\frac{1}{\sqrt{n_y}}\right),
\end{align*}
where the last equation holds when $E[(\hat m_i-m_i)^4]=o(n_y^{-1})$
and $E[(\hat h_i-h_i)^4]=(n_z^{-1})$. Similarly, we can apply the same arguments to the other two terms. Thus, it follows that
\begin{align*}
	D_{41}=o_p\left(\frac{1}{\sqrt{n_y}}\right).
\end{align*}
Next, we turn to the term $D_{42}$. Note that
\begin{align*}
	D_{42}=&\  \frac{1}{  n_y}\sum_{i\in\D_{2y}} g'_{1i}(m_i-\hat m_i)(\hat g_{2i}-\bar g_{2})= \frac{1}{  n_y}\sum_{i\in\D_{2y}} g'_{1i}(m_i-\hat m_i)(\hat g_{2i}-  g_{2i})\\
	&+ \frac{1}{  n_y}\sum_{i\in\D_{2y}}
	g'_{1i}(m_i-\hat m_i)(  g_{2i}-E g_{2})+ ( E g_{2}-\bar g_{2})\frac{1}{  n_y}\sum_{i\in\D_{2y}}
	g'_{1i}(m_i-\hat m_i)=:\sum_{i=1}^3D_{42i}.
\end{align*}
For the first term $D_{421}$,
\begin{align*}
	|D_{421}|&\leq \frac{1}{  n_y}\sum_{i\in\D_{2y}}
	|g'_{1i}(m_i-\hat m_i)(\hat g_{2i}-  g_{2i})|\\
	&\leq \left[\frac{1}{  n_y}\sum_{i\in\D_{2y}}
	(g'_{1i})^2(m_i-\hat m_i)^2\right]^{1/2}\left[\frac{1}{  n_y}\sum_{i\in\D_{2y}}(\hat g_{2i}-  g_{2i})^2\right]^{1/2}\\
	&\lesssim \left[\frac{1}{  n_y}\sum_{i\in\D_{2y}}
	(m_i-\hat m_i)^2\right]^{1/2}\left[\frac{1}{  n_y}\sum_{i\in\D_{2y}}(\hat h_{i}-  h_{i})^2\right]^{1/2}.
\end{align*}
Note that under condition \ref{condition7}, we have $E[ (\hat m_i-m_i)^2]=o(n_y^{-1/2})$ and $E[ (\hat h_i-h_i)^2]=o(n_z^{-1/2})$. It follows that
\begin{align*}
	D_{421}=o_p\left(\frac{1}{n_y^{1/4}n_z^{1/4}} \right)=o_p\left(\frac{1}{\sqrt{n_y}}+\frac{1}{\sqrt {n_z}}\right).
\end{align*}
Similarly, based on $E g_{2}-\bar g_{2}=o_p(n_z^{-1/4})$ and $E[(\hat m_i-m_i)^2]=o(n_y^{-1/2})$, it can be easily obtained that
\begin{align*}
	D_{423}=o_p\left(\frac{1}{\sqrt{n_y}}+\frac{1}{\sqrt {n_z}}\right).
\end{align*}
It follows that
\begin{align*}
	D_{4}= \frac{1}{  n_y}\sum_{i\in\D_{2y}}
	g'_{1i}(m_i-\hat m_i)(  g_{2i}-E g_{2})+ o_p\left(\frac{1}{\sqrt{n_y}}+\frac{1}{\sqrt {n_z}}\right).
\end{align*}

Thus, we can conclude that
\begin{align*}
	I_{n1}& = \frac{1}{  n_y}\sum_{i\in\D_{2y}}
	g'_{1i}\epsilon_i(g_{2i}-Eg_2)+\frac{1}{  n_y}\sum_{i\in\D_{2y}}
	g'_{1i}(m_i-\hat m_i)(  g_{2i}-E g_{2})  +o_p\left(\frac{1}{\sqrt{n_y}}+\frac{1}{\sqrt {n_z}}\right).
\end{align*}

Similarly, we can get
\begin{align*}
	I_{n2} = \frac{1}{  n_z}\sum_{i\in\D_{2z}}
	g'_{2i}\eta(g_{1i}-Eg_1)+\frac{1}{  n_z}\sum_{i\in\D_{2z}}
	g'_{2i}(h_i-\hat h_i)(  g_{1i}-E g_{1})  +o_p\left(\frac{1}{\sqrt{n_y}}+\frac{1}{\sqrt {n_z}}\right).
\end{align*}

For the term $I_{n3}$, we have
\begin{align*}
	I_{n3}&=\frac{1}{ {n}}\sum_{i\in\D_2}
	(\hat g_{2i}-g_{2i}+g_{2i}-Eg_2+Eg_2-\bar{g}_2)(\hat g_{1i}-g_{1i}+g_{1i}-Eg_{1}+Eg_1-\bar{g}_1)\\
	&=\frac{1}{ {n}}\sum_{i\in\D_2}
	(g_{2i}-Eg_2)(g_{1i}-Eg_1)+\frac{1}{ {n}}\sum_{i\in\D_2}
	(\hat g_{2i}-g_{2i})(g_{1i}-Eg_1 )+\frac{1}{ {n}}\sum_{i\in\D_2}
	(\hat g_{1i}-g_{1i})(g_{2i}-Eg_2 )\\
	&\quad +\frac{1}{ {n}}\sum_{i\in\D_2}
	(\hat g_{2i}-g_{2i})(\hat g_{1i}-g_{1i} ) +(Eg_1-\bar{g}_1)\frac{1}{ {n}}\sum_{i\in\D_2}
	(\hat g_{2i}-g_{2i} )+(Eg_1-\bar{g}_1)\frac{1}{ {n}}\sum_{i\in\D_2}
	( g_{2i}-E{g}_2)\\
	&\quad +(Eg_1-\bar{g}_1)
	( Eg_2-\bar {g}_2)+(Eg_2-\bar{g}_2)\frac{1}{ {n}}\sum_{i\in\D_2}
	(\hat g_{1i}-g_{1i} )+(Eg_2-\bar{g}_2)\frac{1}{ {n}}\sum_{i\in\D_2}
	( g_{1i}-E{g}_1)=:\sum_{i=1}^9I_{n3i}.
\end{align*}
For the term $I_{n34}$,
\begin{align*}
	|I_{n34}|&\leq  \frac{1}{ {n}}\sum_{i\in\D_2}
	|(\hat g_{2i}-g_{2i})(\hat g_{1i}-g_{1i} )|\leq \left[\frac{1}{  n}\sum_{i\in\D_{2}}
	(\hat g_{2i}-g_{2i})^2\right]^{1/2}\left[\frac{1}{  n}\sum_{i\in\D_{2}}(\hat g_{1i}-g_{1i})^2\right]^{1/2}\\
	&\lesssim \left[\frac{1}{  n}\sum_{i\in\D_{2}}
	(\hat h_{i}-h_{i})^2\right]^{1/2}\left[\frac{1}{  n}\sum_{i\in\D_{2}}(\hat m_{i}-m_{i})^2\right]^{1/2}=o_p\left(\frac{1}{n_y^{1/4}n_z^{1/4}} \right)=o_p\left(\frac{1}{\sqrt{n_y}}+\frac{1}{\sqrt {n_z}}\right).
\end{align*}
Similarly, it can be easy to show that
\begin{align*}
	\frac{1}{ {n}}\sum_{i\in\D_2}
	(\hat g_{1i}-g_{1i} )=o_p(n_y^{-1/4}),~\mbox{and}~\frac{1}{ {n}}\sum_{i\in\D_2}
	(\hat g_{2i}-g_{2i} )=o_p(n_z^{-1/4}).
\end{align*}

Also, based on the facts that $Eg_1-\bar g_1=o_p(n_y^{-1/4})$ and $Eg_2-\bar g_2=o_p(n_z^{-1/4})$, it can be easily to show that
\begin{align*}
	I_{n35}+I_{n37}+I_{n38}=&\ (Eg_1-\bar{g}_1)\frac{1}{ {n}}\sum_{i\in\D_2}
	(\hat g_{2i}-g_{2i} )+ (Eg_1-\bar{g}_1)
	( Eg_2-\bar {g}_2)\\
	&+(Eg_2-\bar{g}_2)\frac{1}{ {n}}\sum_{i\in\D_2}
	(\hat g_{1i}-g_{1i} ) =o_p\left(\frac{1}{\sqrt{n_y}}+\frac{1}{\sqrt {n_z}}\right).
\end{align*}
Further, note that
\begin{align*}
	\frac{1}{ {n}}\sum_{i\in\D_2}
	( g_{2i}-E{g}_2)=O_p(n^{-1/2}),~\mbox{and}~\frac{1}{ {n}}\sum_{i\in\D_2}
	( g_{1i}-E{g}_1)=O_p(n^{-1/2}),
\end{align*}
which implies that
\begin{align*}
	I_{n36}+I_{n39}= (Eg_1-\bar{g}_1)\frac{1}{ {n}}\sum_{i\in\D_2}
	( g_{2i}-E{g}_2)+  (Eg_2-\bar{g}_2)\frac{1}{ {n}}\sum_{i\in\D_2}
	( g_{1i}-E{g}_1)=o_p\left(\frac{1}{\sqrt{n_y}}+\frac{1}{\sqrt {n_z}}\right).
\end{align*}

Note that by the Taylor series expansion
\begin{align*}
	I_{n32}&=\frac{1}{ {n}}\sum_{i\in\D_2}
	(\hat g_{2i}-g_{2i})(g_{1i}-Eg_1 )\\
	&= \frac{1}{ {n}}\sum_{i\in\D_2}
	g'_2( h_i)(\hat h_{i}-h_{i})(g_{1i}-Eg_1 )+O_p\left(\frac{1}{ {n}}\sum_{i\in\D_2}
	(\hat h_{i}-h_{i})^2(g_{1i}-Eg_1 )\right)\\
	&= \frac{1}{ {n}}\sum_{i\in\D_2}
	g'_{2i}(\hat h_{i}-h_{i})(g_{1i}-Eg_1 )+o_p\left(\frac{1}{\sqrt n_z}\right).
\end{align*}
The last equation holds when $E[(\hat h_i-h_i)^4]=o(n_z^{-1})$ and condition \ref{condition6} are satisfied.
Similarly, we can get
\begin{align*}
	I_{n33} &= \frac{1}{ {n}}\sum_{i\in\D_2}
	(\hat g_{1i}-g_{1i})(g_{2i}-Eg_2 )\\
	&= \frac{1}{ {n}}\sum_{i\in\D_2}
	g'_{1i}(\hat m_{i}-m_{i})(g_{2i}-Eg_2)+O_p\left(\frac{1}{ {n}}\sum_{i\in\D_2}
	(\hat m_{i}-m_{i})^2(g_{2i}-Eg_2 )\right)\\
	&= \frac{1}{ {n}}\sum_{i\in\D_2}
	g'_{1i}(\hat m_{i}-m_{i})(g_{2i}-Eg_2)+o_p\left(\frac{1}{\sqrt n_y}\right).
\end{align*}

Based on the above results, we can conclude that
\begin{align*}
	I_{n}=&\ \frac{1}{  n_y}\sum_{i\in\D_{2y}}
	g'_{1i}\epsilon_i(g_{2i}-Eg_2)+\frac{1}{  n_z}\sum_{i\in\D_{2z}}
	g'_{2i}\eta(g_{1i}-Eg_1)+\frac{1}{ {n}}\sum_{i\in\D_2}
	(g_{2i}-Eg_2)(g_{1i}-Eg_1)\\
	&+\frac{1}{  n_z}\sum_{i\in\D_{2z}}
	g'_{2i}(h_i-\hat h_i)(  g_{1i}-E g_{1}) +\frac{1}{ {n}}\sum_{i\in\D_2}
	g'_{2i}(\hat h_i- h_i)(  g_{1i}-E g_{1})\\
	&+\frac{1}{  n_y}\sum_{i\in\D_{2y}}
	g'_{1i}(m_i-\hat m_i)(  g_{2i}-E g_{2})+\frac{1}{ {n}}\sum_{i\in\D_2}
	g'_{1i}(\hat m_i-m_i)(  g_{2i}-E g_{2}) +o_p\left(\frac{1}{\sqrt{n_y}}+\frac{1}{\sqrt {n_z}}\right)
\end{align*}

Denote $$R_1=\frac{1}{  n_z}\sum_{i\in\D_{2z}}
g'_{2i}(h_i-\hat h_i)(  g_{1i}-E g_{1}) +\frac{1}{ {n}}\sum_{i\in\D_2}
g'_{2i}(\hat h_i- h_i)(  g_{1i}-E g_{1}).$$

Further let $l_i=g'_{2i}(\hat h_i- h_i)(  g_{1i}-E g_{1})$. Note that
\begin{align*}
	R_1=\frac{1}{n}\sum_{i\in\D_{2}}l_i
	-\frac{1}{n_z}\sum_{i\in\D_{2z}}l_i = \frac{n-n_z}{n}\left( \frac{1}{n-n_z}\sum_{i\in\D_{2}/\D_{2z}}l_i
	- \frac{1}{n_z}\sum_{i\in\D_{2z}}l_i\right).
\end{align*}
Similarly as the discussion in the proof of Theorem~\ref{theorem_asy1}, it can be obtained that   $E(R_1)=0$, and
\begin{align*}
	\mbox{Var}(R_1) \leq \frac{2(n-n_z)^2}{n^2}\left(\frac{1}{n-n_z}+\frac{1}{n_z}\right)E(l^2_i).
\end{align*}
Note that $El_i^2=o(1)$, we then  derive that
\begin{align*}
	R_1=\frac{1}{  n_z}\sum_{i\in\D_{2z}}
	g'_{2i}(h_i-\hat h_i)(  g_{1i}-E g_{1}) +\frac{1}{ {n}}\sum_{i\in\D_2}
	g'_{2i}(\hat h_i- h_i)(  g_{1i}-E g_{1})=o_p\left( \frac{1}{\sqrt{n_z}}\right).
\end{align*}
Similarly, we can show that
\begin{align*}
	\frac{1}{  n_y}\sum_{i\in\D_{2y}}
	g'_{1i}(m_i-\hat m_i)(  g_{2i}-E g_{2})+\frac{1}{ {n}}\sum_{i\in\D_2}
	g'_{1i}(\hat m_i-m_i)(  g_{2i}-E g_{2})=o_p\left( \frac{1}{\sqrt{n_y}}\right).
\end{align*}

Thus, we have
\begin{align*}
	I_{n}=&\ \frac{1}{  n_y}\sum_{i\in\D_{2y}}
	g'_{1i}\epsilon_i(g_{2i}-Eg_2)+\frac{1}{  n_z}\sum_{i\in\D_{2z}}
	g'_{2i}\eta(g_{1i}-Eg_1)\\
	&+\frac{1}{ {n}}\sum_{i\in\D_2}
	(g_{2i}-Eg_2)(g_{1i}-Eg_1)  +o_p\left(\frac{1}{\sqrt{n_y}}+\frac{1}{\sqrt {n_z}}\right).
\end{align*}

Similarly for $I_n'$, we have:
\begin{align*}
	I'_{n}=&\ \frac{1}{  n_y}\sum_{i\in\D_{1y}}
	g'_{1i}\epsilon_i(g_{2i}-Eg_2)+\frac{1}{  n_z}\sum_{i\in\D_{1z}}
	g'_{2i}\eta(g_{1i}-Eg_1)\\
	&+\frac{1}{ {n}}\sum_{i\in\D_1}
	(g_{2i}-Eg_2)(g_{1i}-Eg_1)  +o_p\left(\frac{1}{\sqrt{n_y}}+\frac{1}{\sqrt {n_z}}\right).
\end{align*}

Based on the above results, we can conclude that
\begin{align*}
	I_n^*=&\ \frac{1}{N_y}\sum_{i\in\D_{y}}g'_{1i}\epsilon_i(g_{2i}-Eg_2)+\frac{1}{N_z}\sum_{i\in\D_{z}}g'_{2i}\eta(g_{1i}-Eg_1)+\frac{1}{N}\sum_{i\in\D}(g_{2i}-Eg_2)(g_{1i}-Eg_1) \\
	&+o_p\left(\frac{1}{\sqrt{N_y}}+\frac{1}{\sqrt{N_z}}\right).
\end{align*}

Given the asymptotic results of $I^*_{n}$, we can get the results of $\rho^*_n$ by using similar arguments of the proof of Theorem \ref{theorem_asyp} and thus omitted here.

\bibliographystyle{apalike} % Style BST file (imsart-number.bst or imsart-nameyear.bst)
\renewcommand\refname{References}
\bibliography{bibliography}       % Bibliography file (usually '*.bib')

\end{document}